\documentstyle[aps,prb,twocolumn,floats,psfig]{revtex}

\draft
\begin{document}
\twocolumn[\hsize\textwidth\columnwidth\hsize\csname@twocolumnfalse%
\endcsname

\title{Lattice anisotropy as microscopic origin of static stripes in 
cuprates}

\author{B. Normand and A. P. Kampf}

\address{Theoretische Physik III, Elektronische Korrelationen und 
Magnetismus, Institut f\"ur Physik, \\ Universit\"at Augsburg, D-86135 
Augsburg, Germany.}

\date{\today}

\maketitle

\begin{abstract}

Structural distortions in cuprate materials offer a microscopic origin 
for anisotropies in electron transport in the basal plane. Using a 
real-space Hartree-Fock approach, we consider the ground states of the 
anisotropic Hubbard ($t_x \ne t_y$) and $t$-$J$ ($t_x \ne t_y$, $J_x 
\ne J_y$) models. Symmetrical but inhomogeneous (``polaronic'') charge 
structures in the isotropic models are altered even by rather small 
anisotropies to one-dimensional, stripe-like features. We find two 
distinct types of stripe, namely uniformly filled, antiphase 
domain walls and non-uniform, half-filled, in-phase ones. We 
characterize their properties, energies and dependence on the model 
parameters, including filling and anisotropy in $t$ (and $J$). We 
discuss the connections among these results, other theoretical studies 
and experimental observation. 

\end{abstract}
\bigskip
]

\section{Introduction}

The issue of inhomogeneous charge and spin order in cuprate systems, or 
``stripe phases'', has become one of the most controversial issues in the 
debate over high-temperature superconductivity. Experimentally, the 
evidence for static stripe phases in a variety of materials is 
incontrovertible: neutron diffraction studies of the Nd-doped LSCO system 
La$_{2-x-y}$Nd$_y$Sr$_x$CuO$_4$\cite{rtsanu,rtainun,rtaimnu} show over the 
superconducting range of $x$ a charge modulation which has been interpreted 
as arising from a system of charged, alternating (1,0) and (0,1) domain 
walls separating antiferromagnetic (AF) regions where no holes are present. 
A similar phenomenon has been reported more recently in very low-doped LSCO 
without Nd, although in this case the stripes are diagonal and the system 
insulating.\cite{rwsehuybklgl,rwbkleglfyehs} The variation of this 
inhomogeneous charge structure with hole doping $x$ has been characterized 
from the incommensurate magnetic diffraction peaks due to the antiphase 
nature of the domain walls.\cite{rtaimnu,rylesbk} Although dynamical stripes 
(below) have been proposed as a pairing mechanism,\cite{rekzf} also beyond 
dispute is that static stripe formation is inimical to superconductivity, 
as observed both from the individual, possibly coexisting, order parameters 
at fixed Nd content $y$,\cite{rtaimnu} and more broadly from the evolution 
of $T_c$ with $y$.\cite{rbbfk}

Somewhat less unanimity is achieved concerning the existence and role 
of dynamical stripes, by which is meant charge fluctuations of the same 
incommensurate type. In contrast to early proposals based on the shape 
of the Fermi surface,\cite{rlkls,rlzam} such fluctuations are now the 
favored interpretation of the incommensurate peaks observed by inelastic 
neutron scattering in both LSCO\cite{rmam,rcammhcfcm,rylesbk} and 
YBCO\cite{rmdhapd,rmd} systems. Nuclear Quadrupole Resonance (NQR) 
measurements\cite{rhsti,rshci,riutngls} suggest that charge-ordering 
features are ubiquitous, in materials both with and without Nd, on the 
short timescales probed by this technique. Indirect evidence for dynamical 
stripes has also been inferred from the suppression of $T_c$ 
on impurity doping, by the argument that pinning effects lead to static 
stripe formation (see Ref.~\onlinecite{rmcb}, and references therein). 

On the theoretical side, indications for stripe phases predate their 
experimental observation by some 6 years. Initial results from the 
real-space Hartree-Fock approach applied to a three-band Hubbard 
model\cite{rzg} were followed by similar studies in a selection of 
systems,\cite{rpr,rs,rkmnf,rgl,rvllgb,rvr} and by proposals based on 
frustrated phase separation.\cite{rek} A dramatic increase in this 
activity since the discovery of stripes in nickelates and cuprates has 
seen charge inhomogeneities investigated further in Hubbard- and 
$t$-$J$-based models by the Hartree-Fock (HF) approximation,\cite{rzo,rmobb}
HF with additional correlations and static phonons,\cite{rgro} the
slave-boson technique,\cite{rscdg} numerical Density Matrix Renormalization 
Group (DMRG) simulations,\cite{rwsl,rwsb,rmgxfd} the Dynamical Mean-Field 
Theory (DMFT),\cite{rflpo} and Exact Diagonalization (ED).\cite{rmgxfd} 
The results of all of these studies imply that one-dimensional (1d) charge 
ordering is a leading instability in this class of models throughout the 
physical parameter regime. Further analyses using the nonlinear sigma 
model,\cite{rcnh} coupled Luttinger liquids,\cite{rcn} and Cluster 
Perturbation Theory (CPT)\cite{rzeah} adopt this striped state at the 
outset and shed additional light on its microscopic nature. 

At the microscopic level, the relationship between structure and 
superconductivity has also been investigated since before the first 
reports of charge-inhomogeneous phases. B\"uchner {\it et al.}\cite{rbbfk} 
noted the structural transformation and loss of superconductivity on 
increasing Nd doping in a series of samples La$_{2-x-y}$Nd$_y$Sr$_x$CuO$_4$, 
and were further able to relate these results to the bond lengths and 
angle distortions in the low-temperature tetragonal (LTT) phase. Such 
analyses have been extended within the LSCO system to cover a variety 
of substituents,\cite{rdwrjhwhrh} and also towards the use of dynamical 
probes to investigate the relation of structure and spin correlations 
(see for example Ref.~\onlinecite{rkrvbhb}). Of particular interest in 
connection with our analysis is the LTT structural transition also 
caused by Eu doping, and the Cu NQR studies\cite{rtbg} of this phase 
which indicate the presence of three inequivalent Cu sites depending 
on the location of the atom within or outside charged stripe structures. 
The microscopic relationship between lattice structure and the stripe 
instability has also been studied recently by detailed experiments 
focusing on bond lengths, or micro-strain, in the CuO polyhedra.\cite{rbbcdos}

In this work we begin with the experimental motivation provided - and 
quantified - by Ref.~\onlinecite{rbbfk}, that real cuprate systems are 
often anisotropic, and that increase of this anisotropy supports static 
stripe formation. Theoretically, this suggests that appropriate starting 
models for the underlying physics of the cuprate planes would be 
anisotropic Hubbard ($t_x \ne t_y$) or $t$-$J$ ($t_x \ne t_y$, $J_x \ne 
J_y$) models.\cite{rksw} To investigate the relationship between this 
anisotropy and 1d, static charge order, we will adopt the most 
straightforward method known to provide initial, qualitative insight, 
and conduct real-space HF studies of anisotropic Hubbard and $t$-$J$ models. 
We aim to illustrate the effects of anisotropies on the static charge and 
spin configurations, and also to elucidate the energetic contributions 
(kinetic {\it vs.} magnetic) which drive the formation of inhomogeneous 
structures. The unrestricted HF technique is by nature suitable only for 
discussing the static properties, as it neglects quantum fluctuations 
which are essential for dynamical properties and restoration of broken 
symmetries. Thus we will not address the contentious issues of dynamical 
stripes or the connection to superconductivity. However, we hope to provide 
a suitable foundation on which to base further considerations, and will 
discuss below some more sophisticated approaches to the anisotropic models 
which may shed light on these questions, as well as on the roles of quantum 
fluctuations and lattice degrees of freedom which are neglected at this level. 

The structure of the paper is as follows. In section II we review the 
experimental observations of structural anisotropies in cuprate materials. 
In Sec. III we give a brief account of the real-space Hartree-Fock method. 
Sec. IV presents results obtained for the Hubbard model with variable 
hopping anisotropy, and a comprehensive investigation of their evolution 
with the other intrinsic and extrinsic system parameters. In Sec. V we 
consider the $t$-$J$ model, provide analogous results where possible for 
anisotropies in both $t$ and $J$, and compare the two models. Sec. VI 
contains a discussion of the results in the context of experiment, some 
consideration of other materials, models and theoretical approaches, and 
a summary of our analysis. 

\section{Structural Distortions in Cuprates}

Before the first observations of stripes in any oxides, doping of LSCO 
by rare-earth elements was found to cause a structural transition from 
the low-temperature orthorhombic (LTO) phase to 
LTT.\cite{rbbcsshbmhw,rchmfach} B\"uchner {\it et al.} investigated the 
suppression of superconductivity by Nd-doping in LSCO,\cite{rbbfk} finding 
a phase diagram for La$_{2-x-y}$Nd$_y$Sr$_x$CuO$_4$ with the LTO phase 
(Fig.~1(a)) for $y < 0.16$, LTT (Fig.~1(b)) with a superconducting ground 
state in the doping range $0.16 < y < 5 x - 0.55$, and LTT but 
non-superconducting for $y > 5 x - 0.55$. For large $x$ the high-temperature 
tetragonal (HTT) phase remains stable down to low temperature, and for 
small $x$ with $y > 0.16$, the structure is Pccn. By combining transport 
measurements with structural refinement, the line $y \simeq 5 x - 0.55$ 
was found to correspond to a critical buckling angle of the CuO$_6$ 
octahedra in the LTT structure, $\Phi_c \simeq 4^{\rm o}$. A further key 
point in support of the close connection between charge order and lattice 
structure is the observation\cite{rtsanu} that the orientation of the 
stripes alternates between (1,0) and (0,1) on proceeding along the 
${\hat c}$-axis, as does the rotation axis of the LTT structural 
distortion (Fig.~1(b)). It is clear that the electronic structure of 
the CuO$_2$ plane is altered strongly by small changes in the lattice 
structure.\cite{rbrz} 

The LTO and LTT phases both involve a distortion of the flat CuO$_2$ 
planes by rotation of the CuO$_6$ structural units. For LTO this rotation 
takes place around a (1,1) diagonal of the CuO$_2$ system, such that all 
the in-plane O atoms of the HTT structure are displaced from the plane. 
For LTT this rotation is around a (0,1) or (1,0) axis, such that only 
$x$-axis (or $y$-axis) O atoms are displaced. Structural analysis 
indicates that the fundamental Cu-O separation is not altered 
significantly by the distortion, which is taken up rather by the 
Cu-O-Cu bond angle, $2 \Phi$. Inspection of Fig.~1 shows that (despite the 
crystallographic terminology, which is based on a unit cell of side 
$\sqrt{2}$ larger than the Cu-Cu distance), this planar buckling affects 
the Cu-O-Cu bonds equally in $x$- and $y$-directions in the LTO phase, 
but unequally in LTT. 

\medskip
\begin{figure}[t!]
\centerline{\psfig{figure=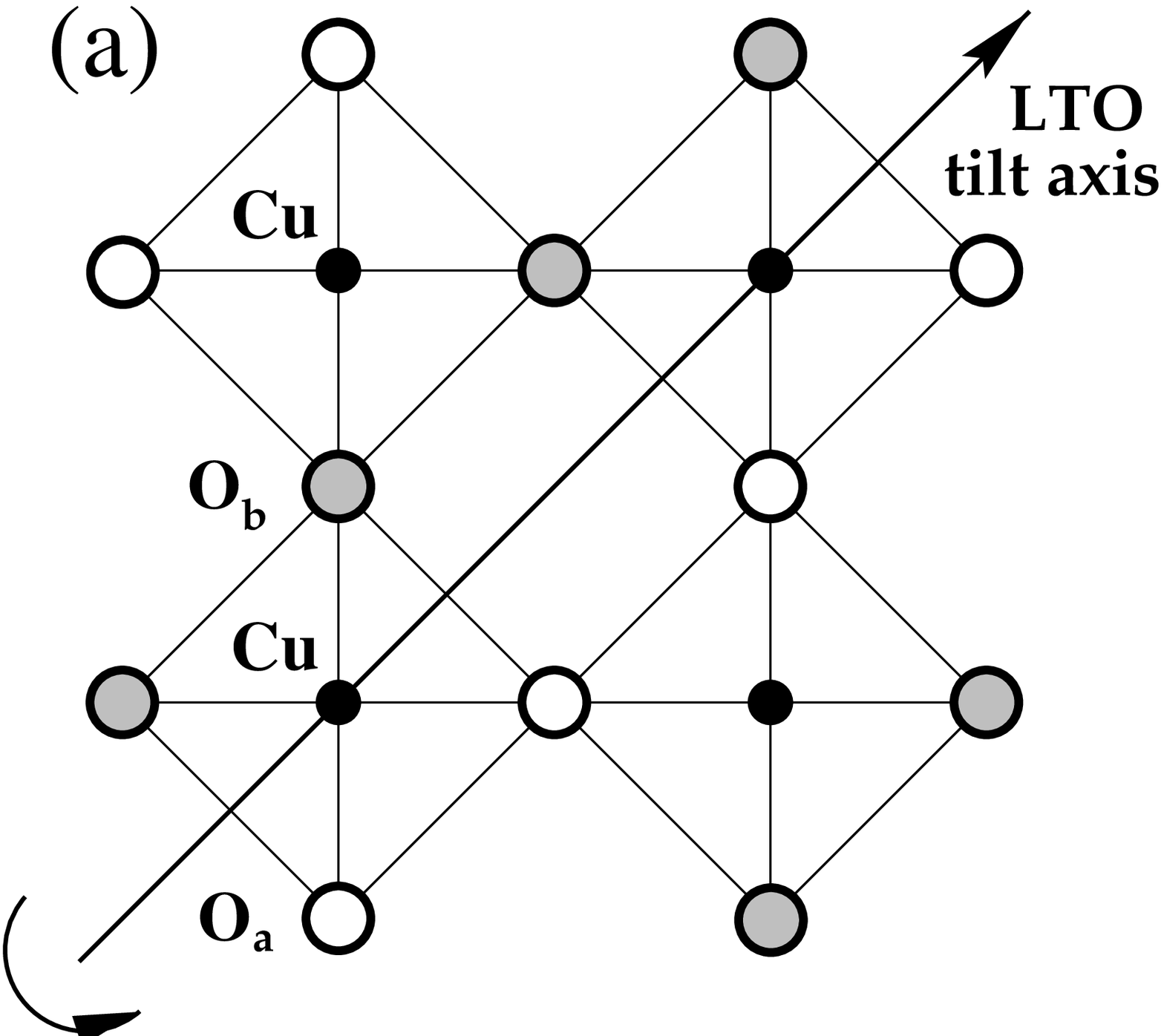,height=5cm,angle=0}}
\centerline{\psfig{figure=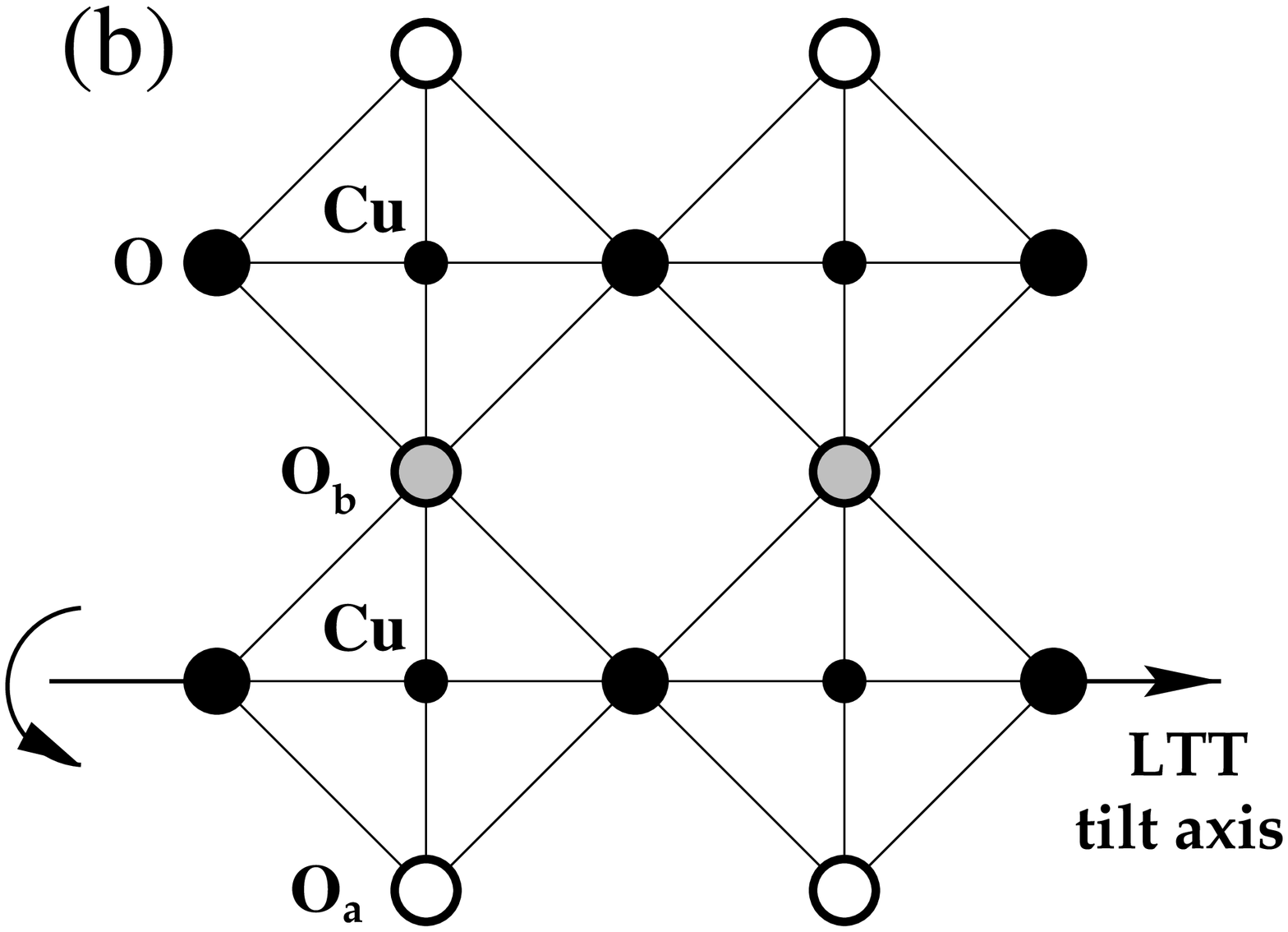,height=5cm,angle=0}}
\medskip
\caption{ Schematic representations of the conformation of the CuO$_2$ 
plane in LTO (a) and LTT (b) lattice structures. Small, black circles 
represent Cu atoms, which remain in the plane in both structures. Large, 
black circles represent O atoms which remain in the plane, white circles 
O atoms which are driven above the plane, and grey circles O atoms which 
are driven below it by the tilting distortion. } 
\end{figure}

This result provides a microscopic origin for in-plane anisotropies in the 
LSCO system. Calculation from first principles of the electron transfer and 
superexchange parameters, $t$ and $J$, indicate that in cuprate materials 
these deviate rather strongly\cite{rej} from the elementary orbital-overlap 
theory.\cite{rh} For a (1,0) LTT anisotropy, one expects an angular 
dependence given approximately by\cite{rksw} 
\begin{equation}
t_y = t_x |\cos (\pi - 2 \Phi)| , \;\; J_y = J_x \cos^2 (\pi - 2 \Phi) .
\label{ehiad}
\end{equation}
For a maximal distortion angle\cite{rbbfk} $\Phi = 5^{\rm o}$, the relative 
anisotropies $|t_x - t_y|/t_x \sim 1.5\%$ and $|J_x - J_y|/J_x \sim 3\%$ 
may appear rather small, but we note that even these small values result 
in the absolute differences $|t_x - t_y|$ and $|J_x - J_y|$ exceeding the 
superconducting $T_c$. The qualitative picture thus emerges of a possible 
microscopic origin for the formation of 1d structures (stripes) as the true
ground state of the weakly anisotropic electronic system, and it is this 
possibility which we wish to investigate. Competition of such a static 
stripe state with a superconducting one, which may be the true ground state 
of the 2d system, would be fully consistent with experimental observation. 
We note in this connection that the issue of bulk coexistence of these two 
states\cite{rtainun} remains open.\cite{rbbfk,rblmhbf}

An important feature of the Nd-doped LSCO system which we have discussed 
is that the LTO-LTT transition and the formation of static stripes, or 
charge order, appear to be decoupled. For $y = 0.4$ and in the optimal 
doping range of $x$, the structural transition occurs at a temperature 
$T_{\rm LT} \sim 80$K, while charge order appears at some $T_{\rm ch} < 
40$K.\cite{rbbfk,riutngls} Although these temperature scales become closer 
for some smaller values of $y$ and $x$, they show opposing trends as 
functions of $x$, and it appears justified to treat the two phenomena 
separately.\cite{rwhb} We will thus consider only the energetics of the 
electronic subsystem, studied on a rigidly distorted lattice. This is 
clearly a simplifying feature in comparison with, for example, highly-doped 
nickelate systems, where stripe formation and structural distortion 
appear to occur together as a single, first-order transition, and any 
energetic considerations would require a comparison between the gain 
in the electronic subsystem with the cost to the lattice. 

In closing this section, we note that another manifestly anisotropic 
cuprate system is YBCO, where the ${\hat b}$-axis chains in the Cu(1)-O 
layer cause a significant difference in the structural parameters of the 
CuO$_2$ plane. The absence of static stripes in YBCO suggests that 
$x$- and $y$-axis transfer integrals may in fact be rather similar 
(to within some critical ratio), or that the bilayer nature may be 
important, or may reflect the fact that the resultant stripe structure 
would be the same in every plane, possibly suffering additional 
penalties in Coulomb energy compared to the case of LTT LSCO, where 
the rotation axis and stripe direction alternate along the $\hat{c}$-axis. 

\section{Real-Space Hartree-Fock Approximation}

This technique has been employed by many authors to gain insight 
into the qualitative physics of the Hubbard model, and we provide only a 
brief overview of the steps involved in the procedure. After the initial 
results of Zaanen and Gunnarsson,\cite{rzg} who found stable, open and 
closed domain walls in a three-band Hubbard model, subsequent 
analyses\cite{rpr,rs,rkmnf,rvllgb} of Hubbard models have verified this 
property in the one-band case, considered the evolution of stripe features 
with the ratio $U/t$ and with filling, and clarified both the filling of 
domain walls and the phase relationship between magnetic regions.\cite{rzo} 
Further refinements of the method have involved the inclusion of local 
correlations and static phonon terms.\cite{rgro}

The one-band Hubbard model is written as 
\begin{equation}
H = \sum_{i, \pm \eta, \sigma } t_{\eta} ( c_{i \pm \eta 
\sigma}^{\dag} c_{i\sigma} + H. c.) + U \sum_i n_{i \uparrow} n_{i 
\downarrow} , 
\label{ehh}
\end{equation}
where $n_{i\sigma} =  c_{i\sigma}^{\dag} c_{i\sigma}$, $\eta$ denotes 
$x,y$, and on an anisotropic lattice $t_x \ne t_y$. Denoting the terms in 
Eq.~(\ref{ehh}) by $H_t$ and $H_U$, in the HF approximation
\begin{eqnarray}
H_U & = & U \sum_i \left[ \langle n_{i \uparrow} \rangle n_{i \downarrow} + 
n_{i \uparrow} \langle n_{i \downarrow} \rangle - \langle n_{i \uparrow} 
\rangle \langle n_{i \downarrow} \rangle \right. \label{ehhf} 
\\ & & \;\; \left. - \langle c_{i\downarrow}^{\dag} 
c_{i\uparrow} \rangle c_{i\uparrow}^{\dag} c_{i\downarrow} - \langle 
c_{i\uparrow}^{\dag} c_{i\downarrow} \rangle c_{i\downarrow}^{\dag} 
c_{i\uparrow} + \langle c_{i\uparrow}^{\dag} c_{i\downarrow} \rangle 
\langle c_{i\downarrow}^{\dag} c_{i\uparrow} \rangle \right] , \nonumber
\end{eqnarray}
where the signs include fermion statistics, precluding the need to form 
Slater determinants of quasiparticle states. We will focus here on collinear 
spin configurations, and not in spiral or vortex solutions.\cite{rvllgb} 
In this case, by choosing the axis of the possibly broken spin symmetry to 
be the $\hat{z}$- axis,\cite{rpr} one may work without loss of generality 
using only the first line of Eq.~(\ref{ehhf}). 

The model is solved in real space on a small cluster of sites $i$. The 
particle number $q_i$ and magnetization components $m_i^{\alpha}$ are 
defined by the expectation values
\begin{mathletters}
\label{ehfev}
\begin{eqnarray} 
q_i & = & \langle c_{i\uparrow}^{\dag} c_{i\uparrow} + c_{i\downarrow}^{\dag} 
c_{i\downarrow} \rangle \\ m_i^z & = & {\textstyle \frac{1}{2}} 
\langle c_{i\uparrow}^{\dag} c_{i\uparrow} - c_{i\downarrow}^{\dag} 
c_{i\downarrow} \rangle \\ m_i^x & = & {\textstyle \frac{1}{2}} \langle 
c_{i\uparrow}^{\dag} c_{i\downarrow} + c_{i\downarrow}^{\dag} c_{i\uparrow} 
\rangle \\ m_i^y & = & - {\textstyle \frac{1}{2}} i \langle 
c_{i\uparrow}^{\dag} c_{i\downarrow} - c_{i\downarrow}^{\dag} c_{i\uparrow} 
\rangle . 
\end{eqnarray} 
\end{mathletters}
Solution proceeds by iteration of the site parameters to self-consistency, 
at which point one has solved all of the mean-field equations contained in 
Eq.~(\ref{ehfev}), subject to the additional constraint on the total particle 
number, which is implemented in the number of filled quasiparticle energy 
levels. This procedure is not identical to finding the global minimum of 
the multi-parameter free-energy surface, in that in principle it may only 
find stationary points, and the user is generally offered no guarantee that 
these correspond to global minima. 

\medskip
\begin{figure}[t!]
\centerline{\psfig{figure=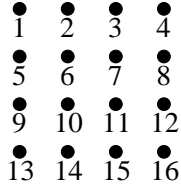,width=2.3cm,angle=0}}
\medskip
\caption{ Numbering scheme for an $l = n \times n$ cluster, illustrated with 
$n = 4$. }
\end{figure}

To solve the self-consistency problem for an $l = n \times n$ cluster requires 
diagonalization of the $2 n^2 \times 2 n^2$ Hamiltonian matrix form of 
Eq.~(\ref{ehh}), 
\begin{equation}
H_{ij} = \left[ \begin{array}{cccccc} {\bf N}_1 & {\bf M}_{12} & 0 & 
{\bf M}_{41}^* & {\bf M}_{15} & \dots \\ {\bf M}_{12}^* & 
{\bf N}_2 & {\bf M}_{23} & 0 & 0 & \dots \\ 0 & {\bf M}_{23}^* & 
{\bf N}_3 & {\bf M}_{34} & 0 & \dots \\ \vdots & \vdots & \vdots & \vdots & 
\vdots & \ddots \end{array} \right] .
\label{ehms}
\end{equation}
Here we have discarded the constant terms in Eq.~(\ref{ehhf}), although 
these are restored in calculating the energy of the self-consistent solutions 
which emerge. The $2 \times 2$ ``diagonal'' and ``off-diagonal'' matrix 
elements are respectively 
\begin{equation}
{\bf N}_i = \left( \begin{array}{cc} {\textstyle \frac{1}{2}} U (q_i - 
2 m_i^z) & - U (m_i^x - i m_i^y) \\ - U (m_i^x + i m_i^y) & {\textstyle 
\frac{1}{2}} U (q_i + 2 m_i^z) \end{array} \right) 
\label{ehhfmd}
\end{equation}
and 
\begin{equation}
{\bf M}_{ij} = \left( \begin{array}{cc} t_{\eta} & 0 \\ 0 & t_{\eta} 
\end{array} \right) ,
\label{ehhfmo}
\end{equation}
in which $\eta = x,y$ for $j = i+1(-n)$, $i+n(-l)$, and the two-component 
structure corresponds to the two spin directions in the $S_z$ basis. 
For the numbering scheme illustrated in Fig.~2 with a $4 \times 4$ cluster, 
the off-diagonal blocks connect $i$ to $i \pm 1$ in the $\hat{x}$-direction, 
and to $i \pm n$ in the $\hat{y}$-direction. Open boundary conditions (BCs) 
are implemented with no forward (backward) connections in $x$ when mod$(i,n) 
= 0$ (mod$(i,n) = n - 1$), and in $y$ when $i > n^2 - n$ ($i \le n$). 
Periodic BCs have forward (backward) connections from $i$ to $i + 1 
- n $ ($i - 1 + n$) in $x$, such as the element ${\bf M}_{41}$ in 
Eq.~(\ref{ehms}), and to $i + n - n^2$ ($i - n + n^2$) in $y$ for the 
same edge sites. Because the Hamiltonian matrix is Hermitian, only half 
of these connections must be set explicitly. For cluster filling $x$, the 
lowest $2 n^2 x$ eigenstates of the diagonal Hamiltonian matrix are filled, 
$q_i$ and ${\bf m}_i$ for all sites $i$ are deduced from the eigenvectors 
representing the new quasiparticles, the new matrix is diagonalized, and 
the process continued. The consistency condition was generally taken to 
have been achieved when the summed squares of the changes in $q_i$ and 
$m_i^{\alpha}$ were less than 10$^{-8}$ per site, although this was 
continued in some cases to the machine precision to ensure that no further 
changes were possible. Convergence was usually achieved for $12 \times 12$ 
systems within 200 iterations using open BCs, and within 1000 iterations 
for periodic BCs, where the iteration procedure could more clearly be 
observed converging on and diverging from a sequence of local minima. 

The $t$-$J$ model has previously been considered in the real-space HF 
formalism only in Ref.~\onlinecite{rvr}. Here we have chosen to study 
the form 
\begin{equation}
H_{t-J} = H_t + H_U + J \left( {\bf S}_i {\bf \cdot S}_j - {\textstyle 
\frac{1}{4}} n_i n_j \right) ,
\label{ehtj}
\end{equation}
with $n_i = n_{i\uparrow} + n_{i\downarrow}$, and $U$ taken to have a 
large, finite value to mimic the projection property of the $t$-$J$ 
model onto only those states with no doubly occupied sites. We will 
have more to say on this approximation in Sec. V. In this case the $J$ 
term admits further decouplings, among which we neglect pairing terms 
when working in a canonical ensemble, and nearest-neighbor spin-flip 
terms because these are generally small. Thus we retain only a finite 
expectation value of the bond hopping term $A_{ij \sigma} = \langle 
c_{i \sigma}^{\dag} c_{j\sigma} \rangle$.\cite{rvr}

The HF decomposition of the additional $J$ term is 
\begin{eqnarray}
H_J & = & - {\textstyle \frac{1}{2}} J \sum_{\langle ij \rangle} 
\left[ \langle n_{i \uparrow} \rangle n_{j \downarrow} + \langle n_{j 
\downarrow} \rangle n_{i \uparrow} - \langle n_{i \uparrow} 
\rangle \langle n_{j \downarrow} \rangle \right. \label{etjhf} \nonumber 
\\ & & \;\;\;\;\;\; + \langle n_{i \downarrow} \rangle n_{j \uparrow} + 
\langle n_{j \uparrow} \rangle n_{i \downarrow} - \langle n_{i \downarrow} 
\rangle \langle n_{j \uparrow} \rangle \nonumber \\ & & \;\;\;\;\;\; 
+ A_{ij \uparrow} c_{j\downarrow}^{\dag} c_{i\uparrow} + A_{ij \uparrow}^* 
c_{i\downarrow}^{\dag} c_{j\uparrow} + A_{ij \uparrow}^* A_{ij \downarrow} 
\\ & & \;\;\;\;\;\; \left. + A_{ij \uparrow} c_{j\downarrow}^{\dag} 
c_{i\uparrow} + A_{ij \uparrow}^* c_{i\downarrow}^{\dag} c_{j\uparrow} 
+ A_{ij \uparrow}^* A_{ij \downarrow} \right] . \nonumber
\end{eqnarray}
The $2n^2 \times 2n^2$ matrix to be diagonalized ({\it cf.}~(\ref{ehms})) 
has 
\begin{equation}
{\bf N}_i = \left( \begin{array}{cc} {\textstyle \frac{1}{2}} [ U (q_i - 
2 m_i^z) - {\tilde J}_{i\downarrow} ] & - U (m_i^x - i m_i^y) \\ - U 
(m_i^x + i m_i^y) & {\textstyle \frac{1}{2}} [ U (q_i + 2 m_i^z) - 
{\tilde J}_{i\uparrow} ] 
\end{array} \right) 
\label{emtjd}
\end{equation}
as the $2 \times 2$ diagonal component, where 
\begin{equation} 
{\tilde J}_{\sigma} = J_x (n_{i+x \sigma} - n_{i-x \sigma}) + J_y (n_{i+y 
\sigma} - n_{i-y \sigma}) , 
\label{ejt}
\end{equation}
and 
\begin{equation}
{\bf M}_{ij} = \left( 
\begin{array}{cc} t_{\eta} - {\textstyle \frac{1}{2}} J_{\eta} A_{\eta 
\downarrow}^* & 0 \\ 0 & t_{\eta} - {\textstyle \frac{1}{2}} J_{\eta} 
A_{\eta \uparrow}^* \end{array} \right) .
\label{ehtjm}
\end{equation}
in the non-zero, off-diagonal blocks of the upper triangle. Further details 
of the iterative calculations in this case are deferred to Sec. V. 

\section{Hubbard Model}

In this section we describe the charge and spin configurations (\ref{ehfev}) 
which emerge as the self-consistent solutions to the Hubbard model (\ref{ehh}) 
in the Hartree-Fock approximation (\ref{ehhf}). We begin by considering the 
isotropic system as a function of the ``intrinsic'' system parameters, the 
ratio $U/t$, the (hole) filling $x$, and the temperature $T$, and make 
contact with the results of previous studies. We then illustrate how these 
results are altered in the presence of hopping anisotropy $t_x \ne t_y$, for 
which we study both the charge and spin distribution and the kinetic and 
magnetic energy components. We comment also on the variation of the solutions 
with ``extrinsic'' parameters, by which is meant cluster size, BCs, and 
commensuration. We now set the energy unit to be $t = 1$. In all figures 
to follow the ${\hat x}$-axis is horizontal and the ${\hat y}$-axis vertical.

\subsection{Variation of \mbox{\bf $U$}}

For small values of $U$ the solutions are homogeneous in charge and have 
no local magnetization. This ``metallic'' phase is the ground state for $U 
\alt 3$, the exact value of $U_c$ depending on the filling. We have not 
analyzed this regime or the metal-insulator transition in any detail, as 
many more accurate techniques exist.\cite{rd} For all values $U \agt 3$, 
the predominant HF solutions are inhomogeneous charge clusters, with 
locally AF spin orientation. In the approximate range $3 \alt U 
\alt 6$, these clusters take the form of closed loops of 
electron-depleted sites, or domain walls enclosing an undoped region 
(Figs.~3(a) and 4(a)). We will call these formations ``corrals'', to 
distinguish them from the situation as $U$ is increased. For $U \agt 6$ 
the charge clusters are better described as a solid region of hole doping, 
which we will call a ``polaron'' (Fig.~3(b)). Our results are rather 
similar to those in Fig.~1 of Verg\'es {\it et al.}\cite{rvllgb} for this 
isotropic situation.

\medskip
\begin{figure}[t!]
{\centerline{Open BCs \qquad\qquad\qquad\qquad\quad Open BCs}}
\medskip
\mbox{\psfig{figure=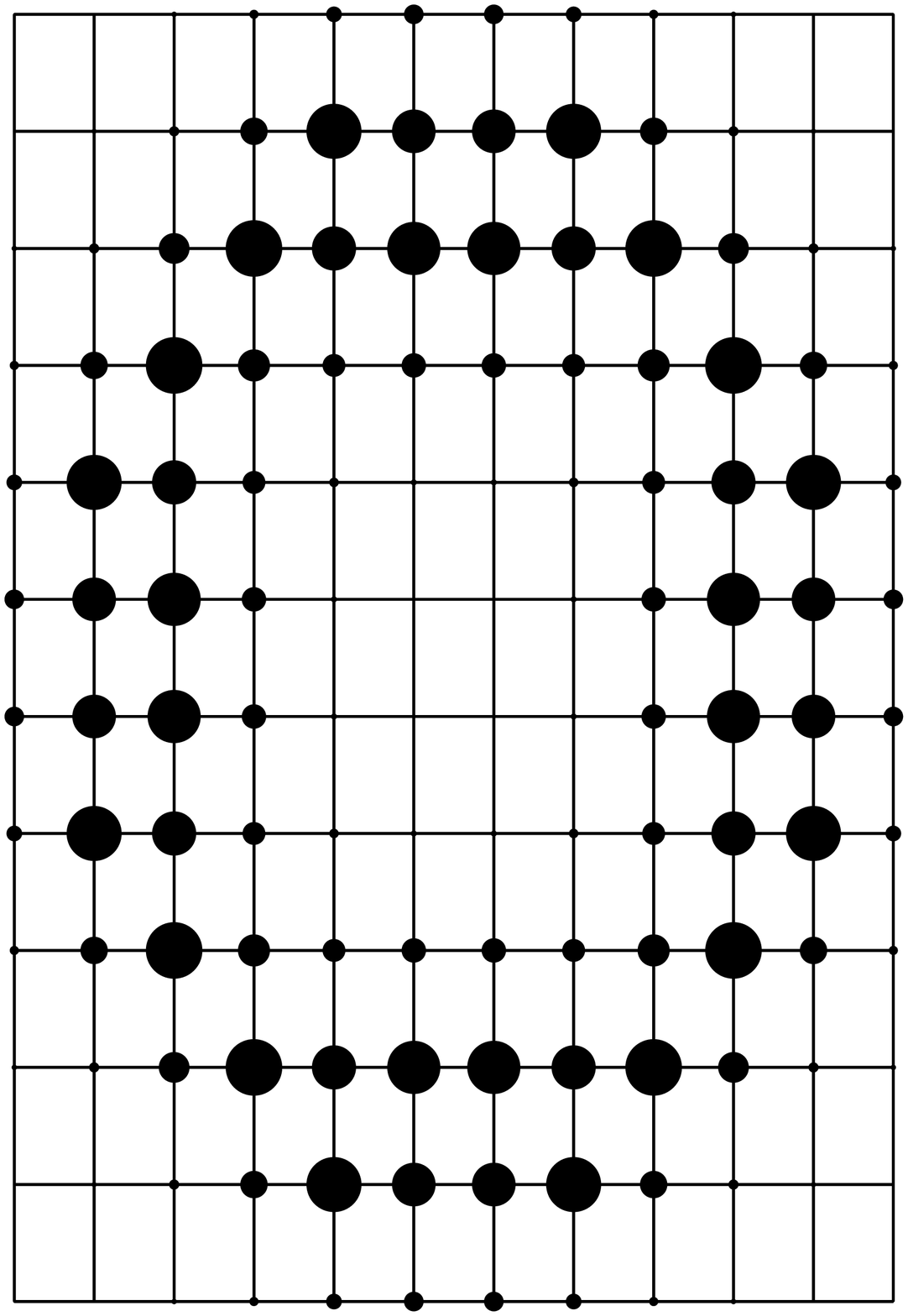,height=5.25cm,angle=0}}
\mbox{\hspace{-0.5cm}\psfig{figure=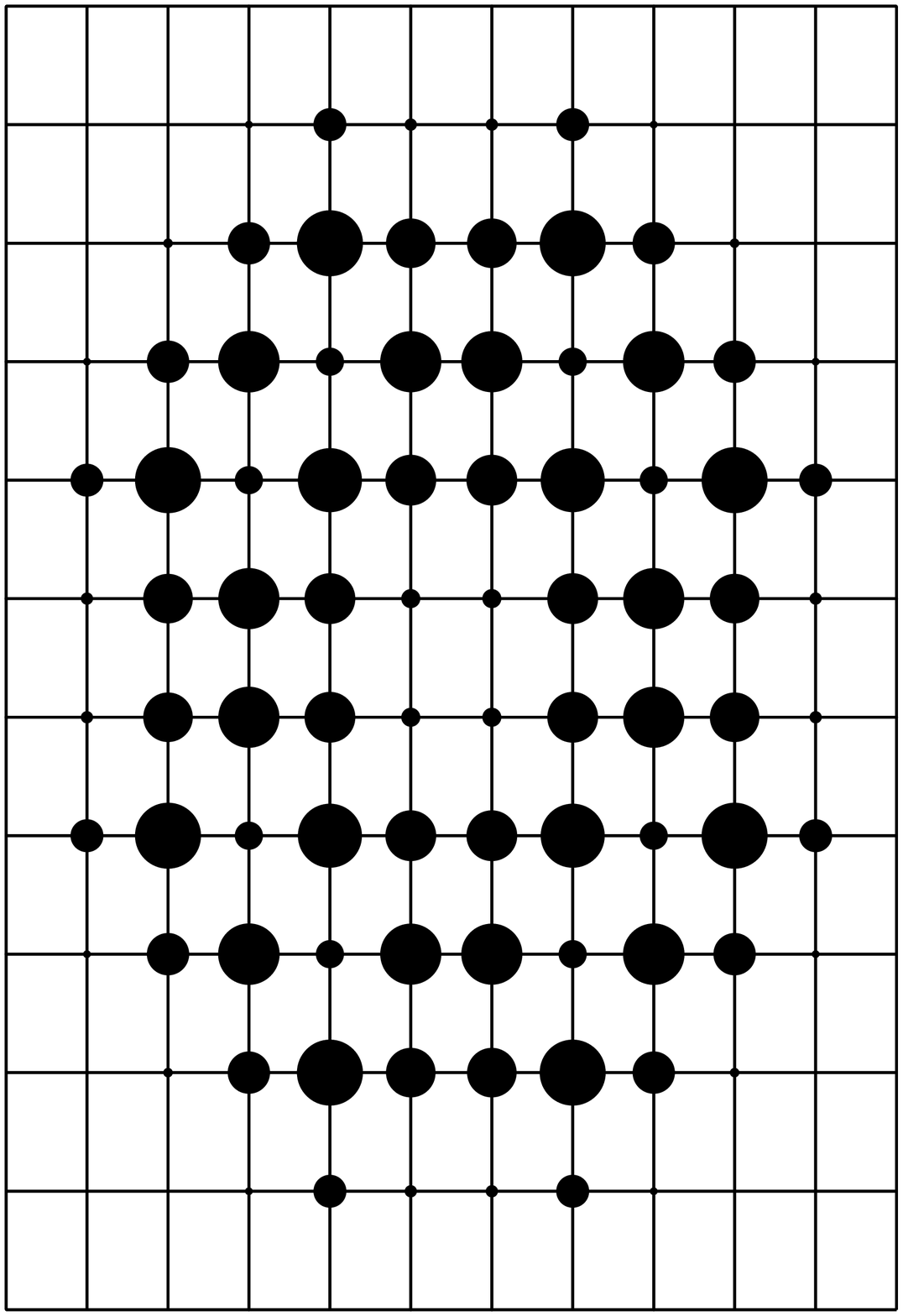,height=5.25cm,angle=0}}
\medskip
{\centerline{(a) \qquad\qquad\qquad\qquad\qquad\qquad (b)}}
\medskip
\mbox{\psfig{figure=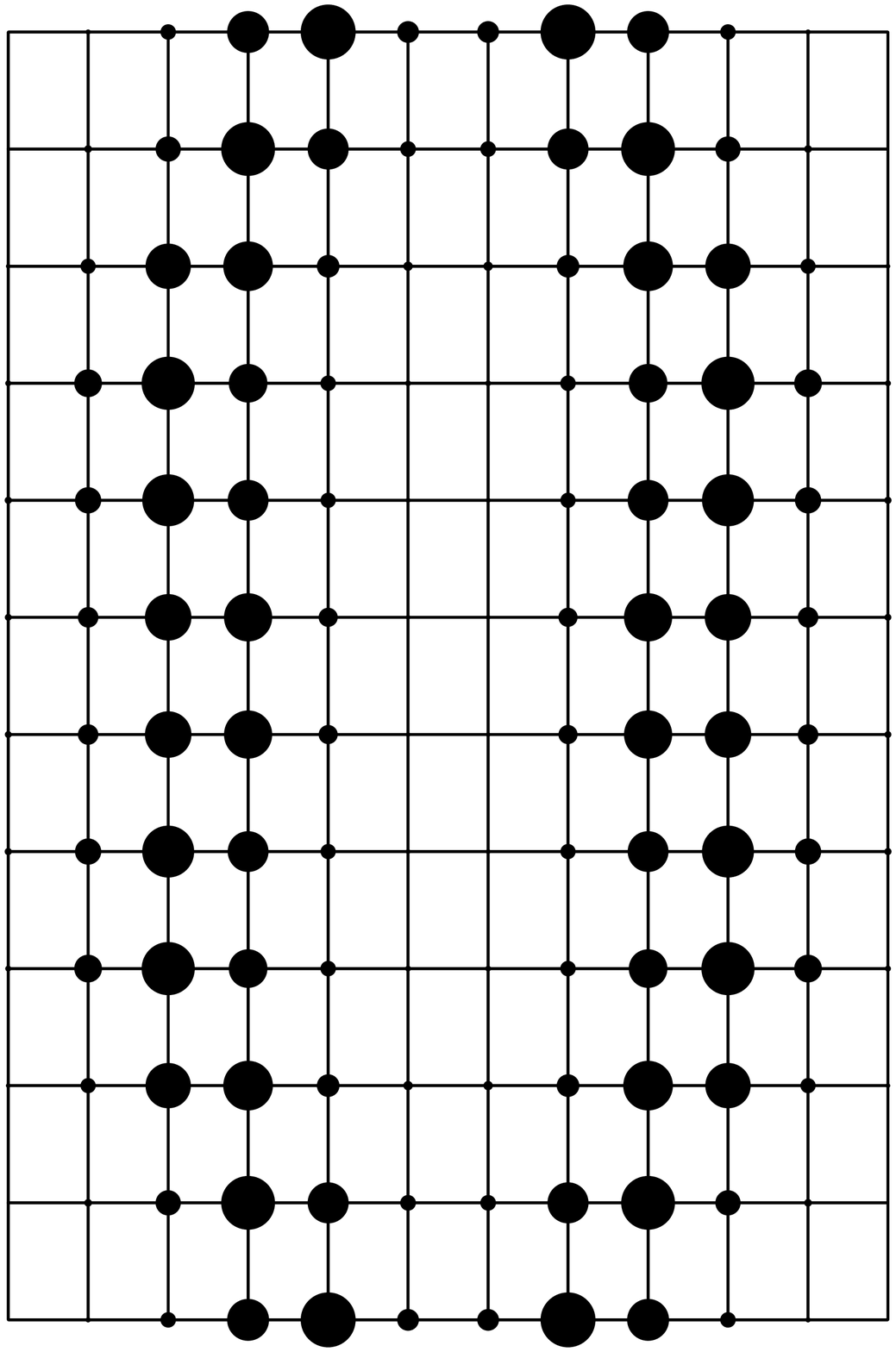,height=5.25cm,angle=0}}
\mbox{\hspace{-0.5cm}\psfig{figure=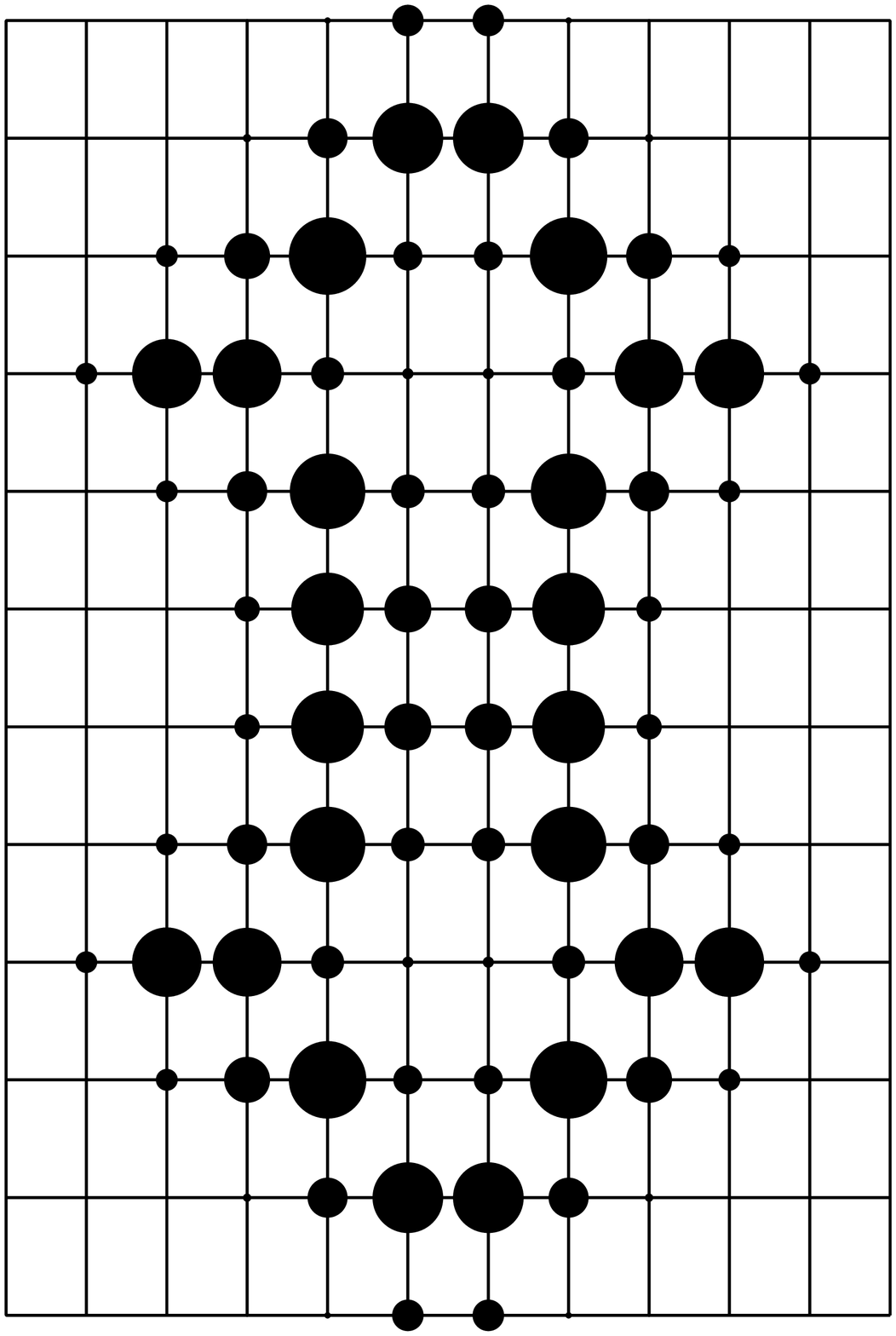,height=5.25cm,angle=0}}
\medskip
{\centerline{(c) \qquad\qquad\qquad\qquad\qquad\qquad (d)}}
\medskip
\mbox{\psfig{figure=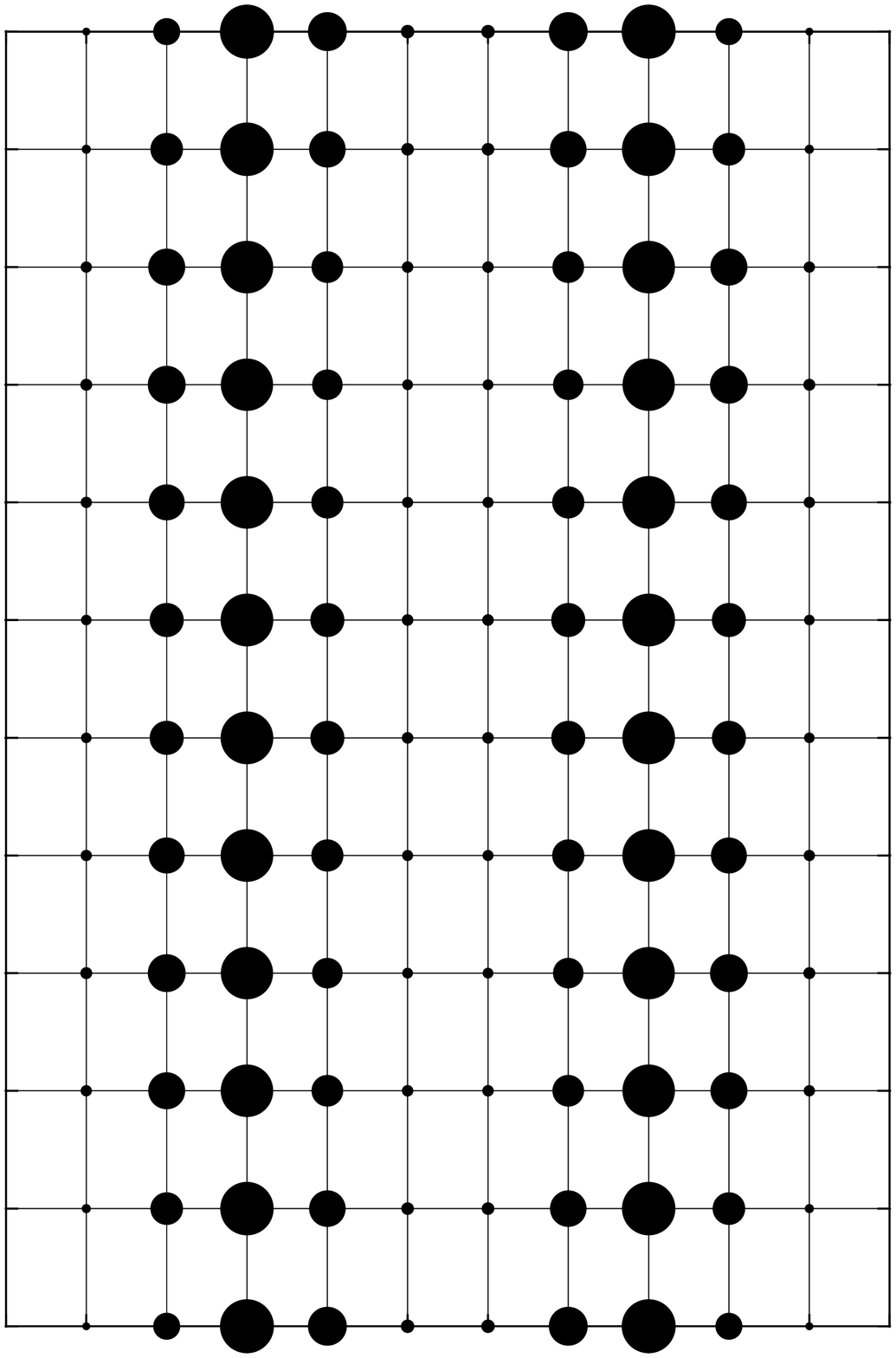,height=5.25cm,angle=0}}
\mbox{\hspace{-0.5cm}\psfig{figure=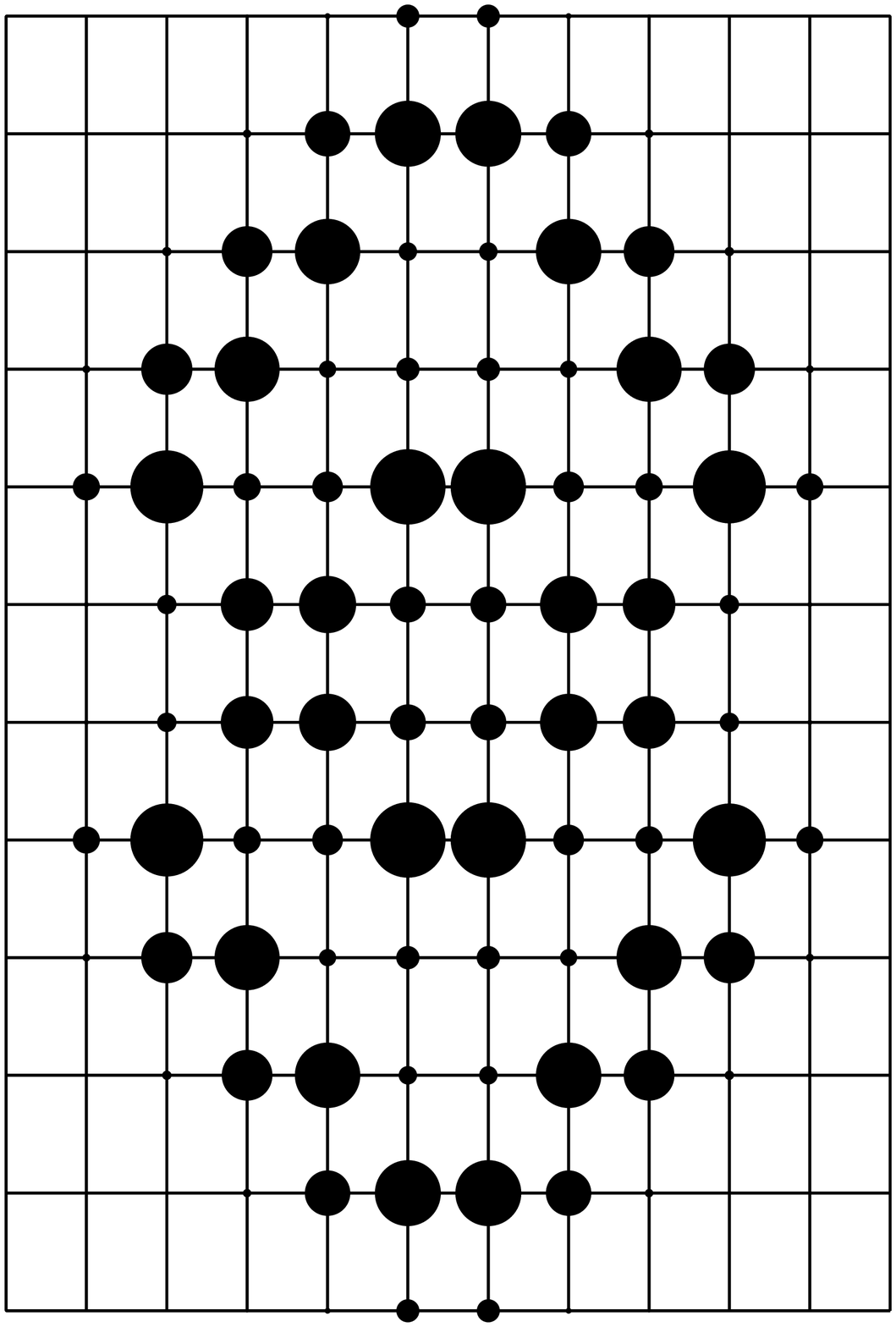,height=5.25cm,angle=0}}
\medskip
{\centerline{(e) \qquad\qquad\qquad\qquad\qquad\qquad (f)}}
\medskip
\caption{Ground-state charge distributions for Hubbard model with hole 
doping $x = 1/6$ on a $12 \times 12$ cluster with open BCs. (a) $U = 5$, 
$t_x = t_y = - 1$. (b) $U = 8$, $t_x = t_y = - 1$. (c) $U = 5$, $t_x = 
- 1.05$, $t_y = - 0.95$. (d) $U = 8$, $t_x = - 1.05$, $t_y = - 0.95$. (e) 
$U = 5$, $t_x = - 1.1$, $t_y = - 0.9$. (f) $U = 8$, $t_x = - 1.1$, $t_y 
= - 0.9$. In these figures the largest circles correspond to $\langle 
n_i \rangle = 0.48$, or 52\% hole doping of the site. }
\end{figure}

\medskip
\begin{figure}[t!]
{\centerline{Open BCs \qquad\qquad\qquad\qquad\quad Periodic BCs}}
\medskip
\mbox{\psfig{figure=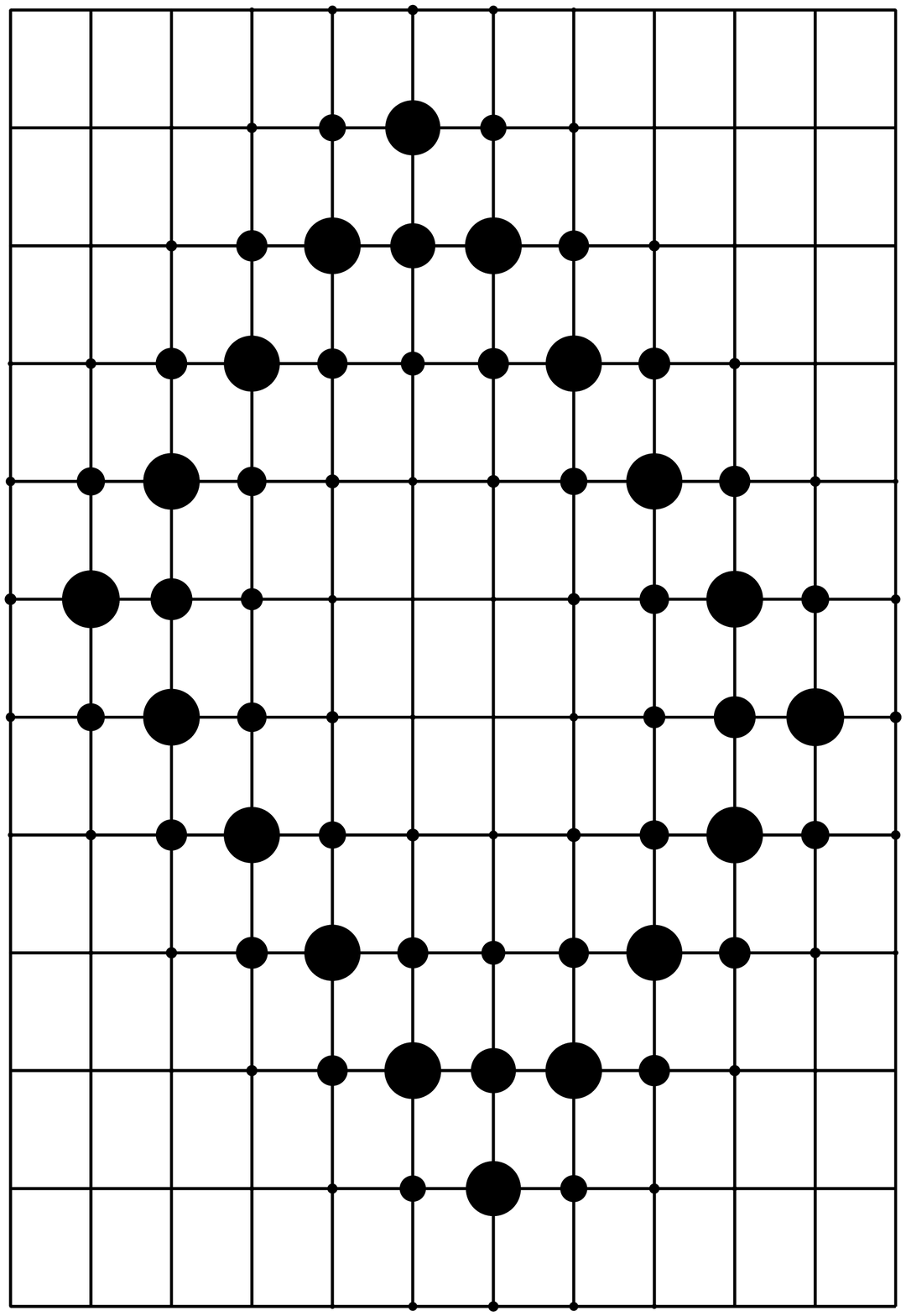,height=5.25cm,angle=0}}
\mbox{\hspace{-0.5cm}\psfig{figure=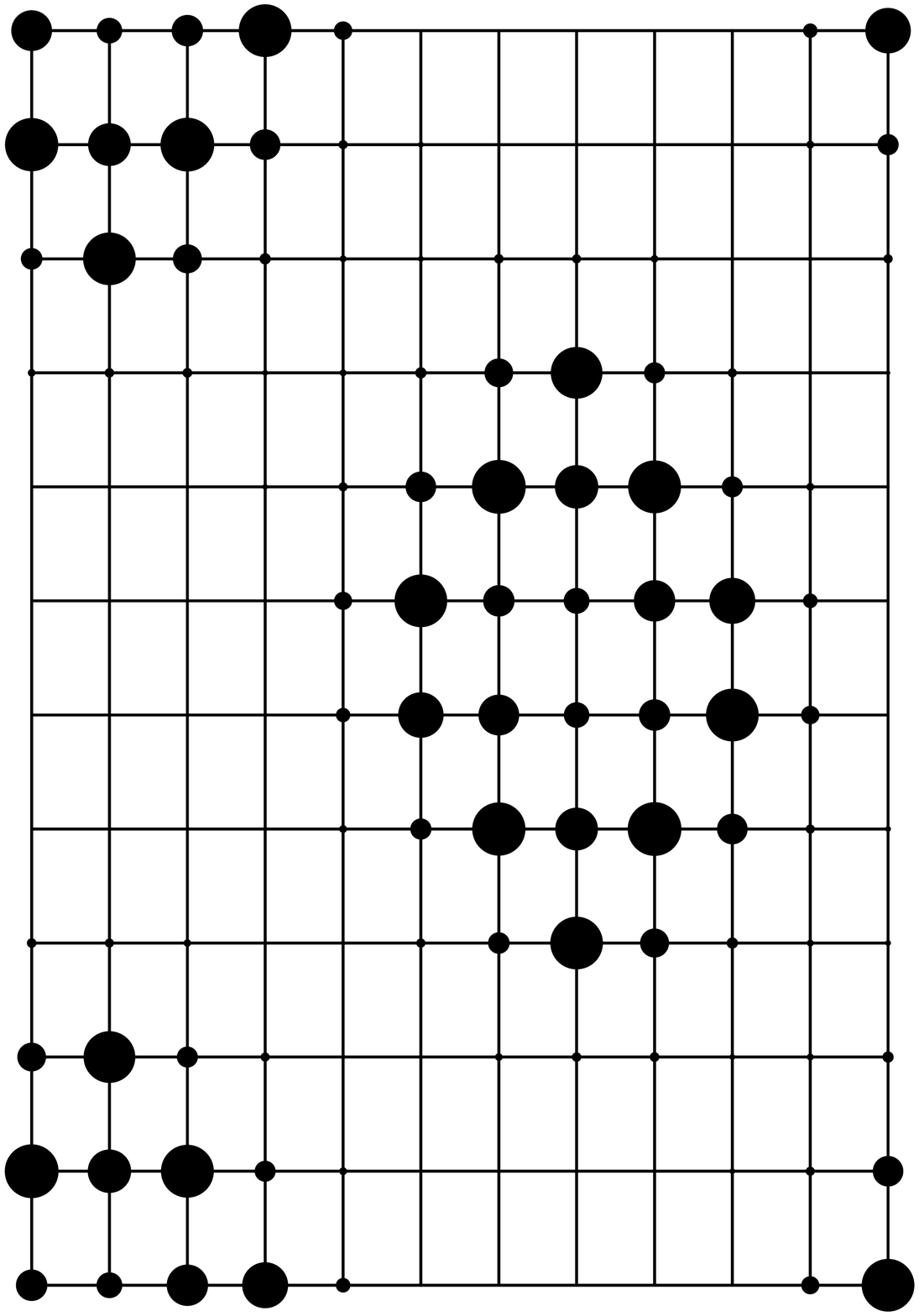,height=5.25cm,angle=0}}
\medskip
{\centerline{(a) \qquad\qquad\qquad\qquad\qquad\qquad (b)}}
\medskip
\mbox{\psfig{figure=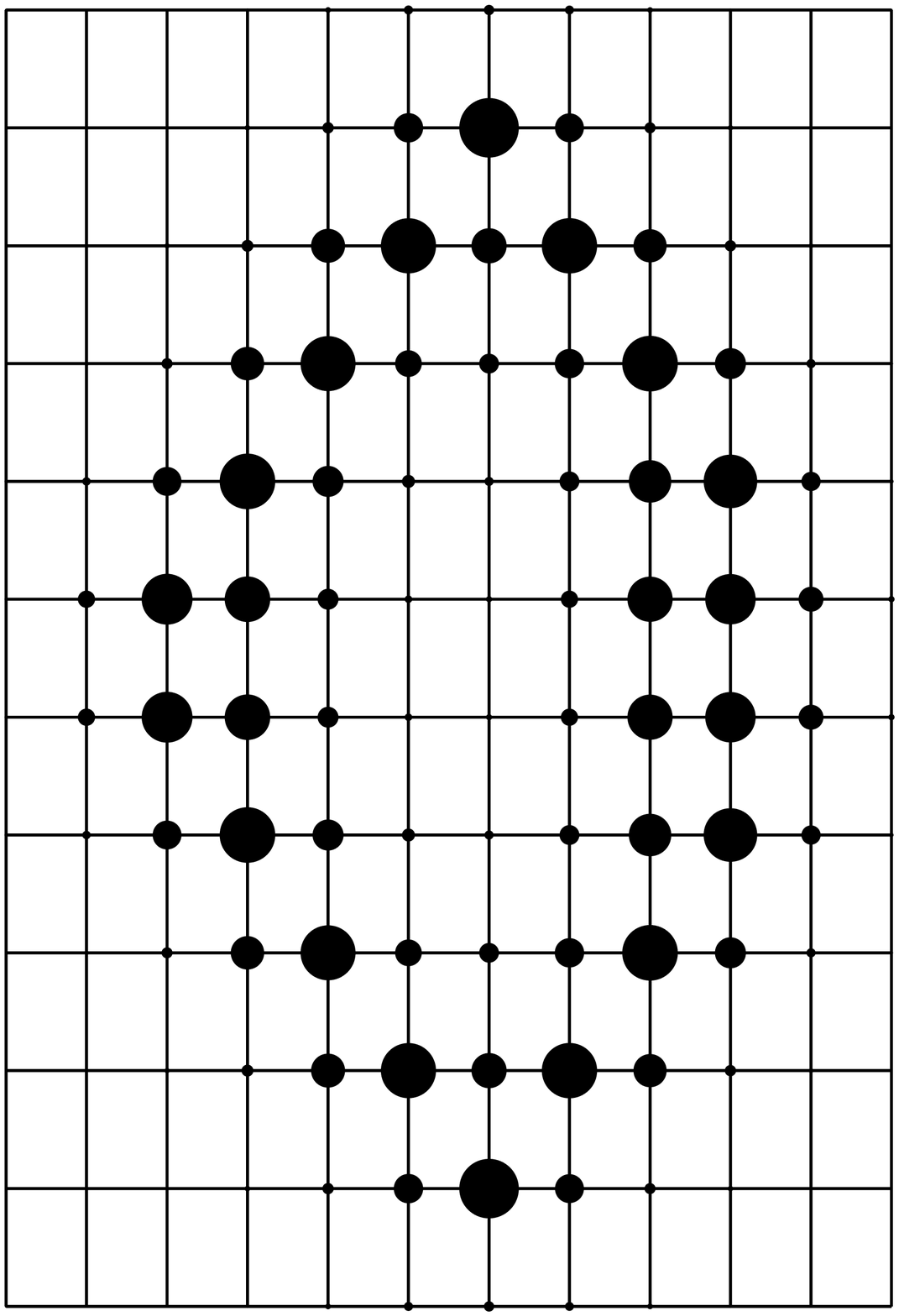,height=5.25cm,angle=0}}
\mbox{\hspace{-0.5cm}\psfig{figure=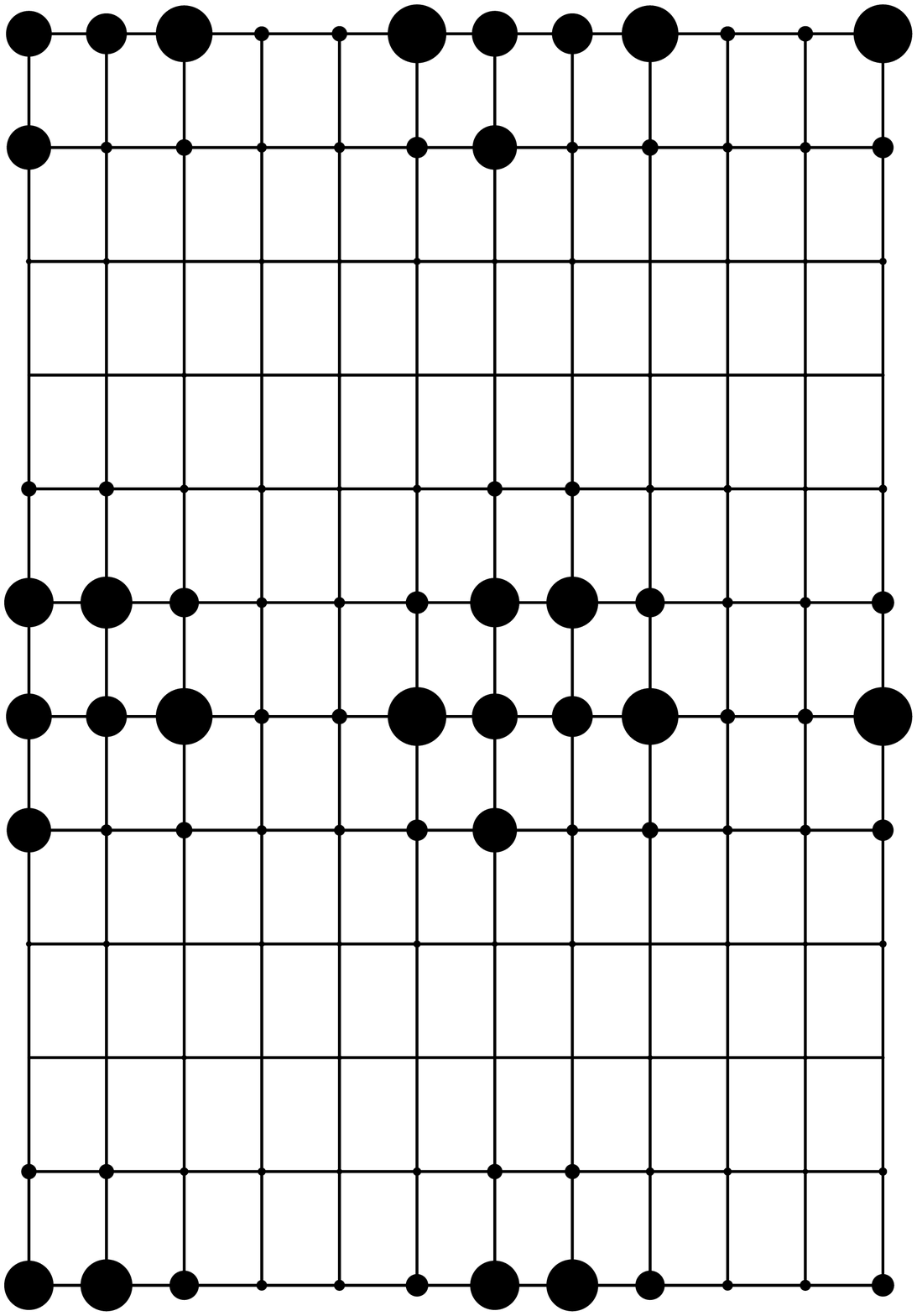,height=5.25cm,angle=0}}
\medskip
{\centerline{(c) \qquad\qquad\qquad\qquad\qquad\qquad (d)}}
\medskip
\mbox{\psfig{figure=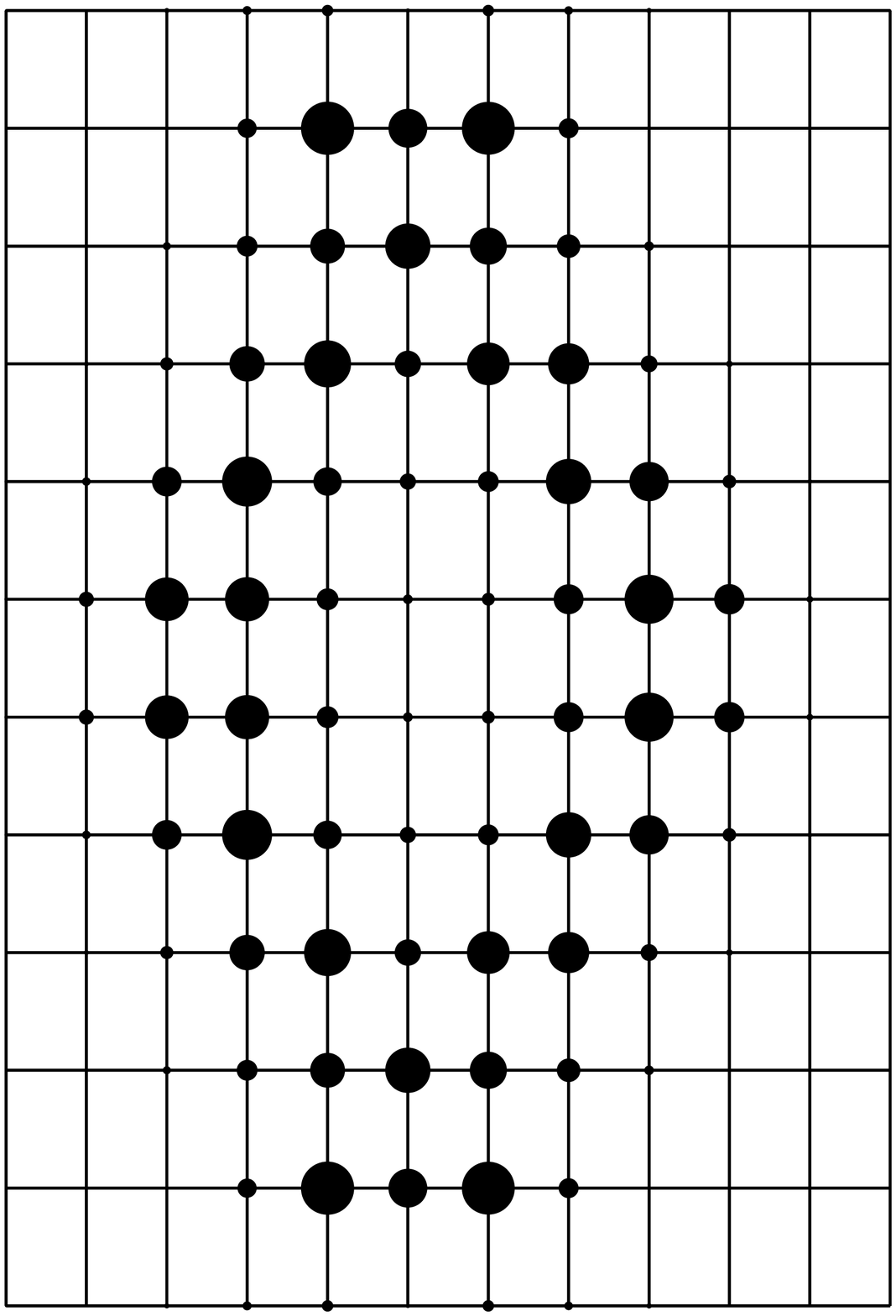,height=5.25cm,angle=0}}
\mbox{\hspace{-0.5cm}\psfig{figure=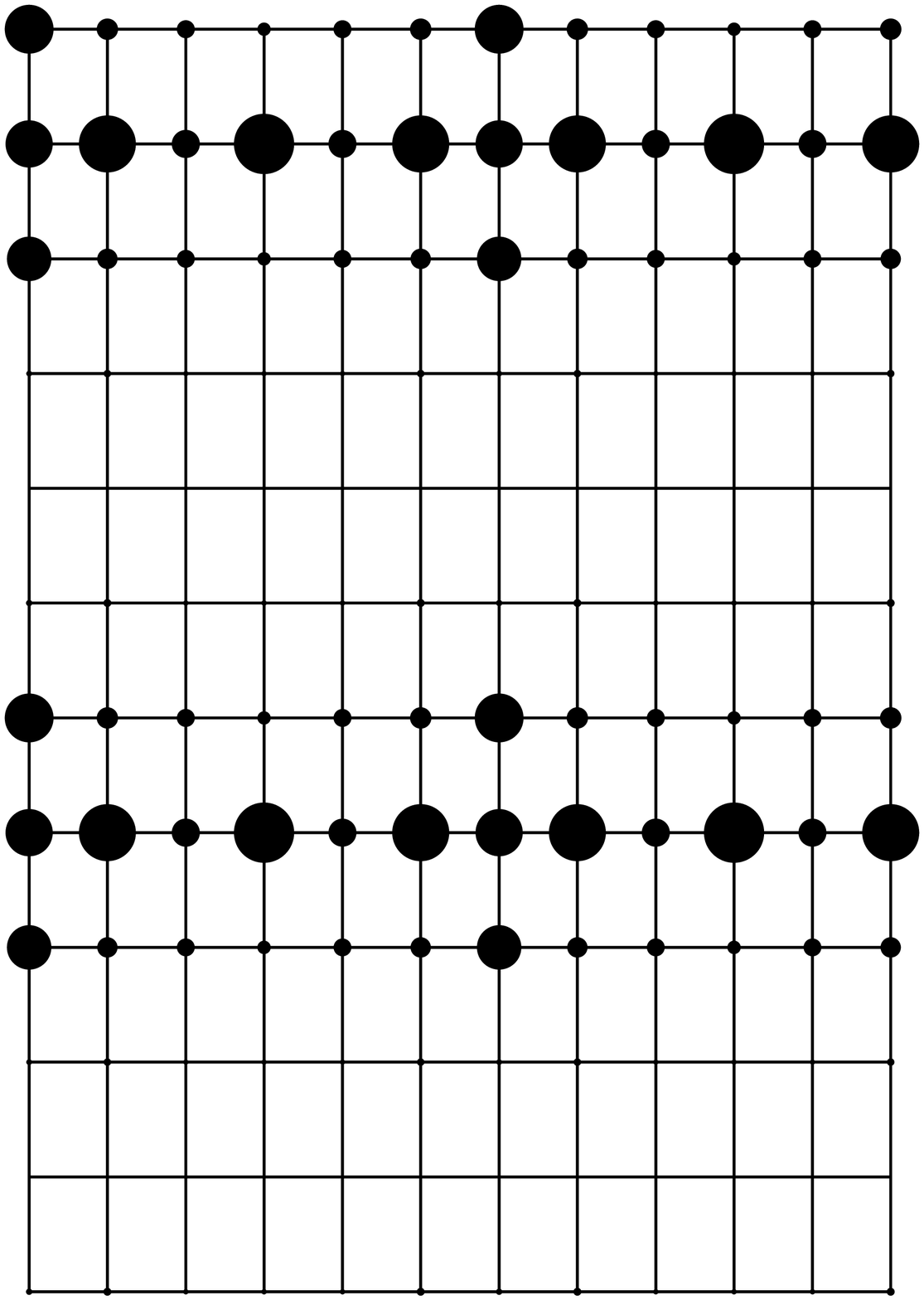,height=5.25cm,angle=0}}
\medskip
{\centerline{(e) \qquad\qquad\qquad\qquad\qquad\qquad (f)}}
\medskip
\caption{Ground-state charge distribution for Hubbard model with hole 
doping $x = 1/8$ and $U = 5$ on a $12 \times 12$ cluster. (a) Open BCs, 
$t_x = t_y = - 1$. (b) Periodic BCs, $t_x = t_y = - 1$. (c) Open BCs, $t_x 
= - 1.05$, $t_y = - 0.95$. (d) Periodic BCs, $t_x = - 1.05$, $t_y = - 0.95$. 
(e) Open BCs, $t_x = - 1.1$, $t_y = - 0.9$. (f) Periodic BCs, $t_x = - 1.1$, 
$t_y = - 0.9$. Charge scale as in Fig.~3. }
\end{figure}

\subsection{Variation of \mbox{\bf $x$}}

As the filling is varied in the isotropic model we note the same general 
features, but with important alterations. With smaller hole doping the 
cluster is less influenced by the system boundaries, and prefers 
(Fig.~4(a)) ``diagonal'' corrals composed of (1,1) 
domain walls, rather than the ``vertical'' type with (1,0) domain walls 
(Fig.~3(a)). This result is quite ubiquitous, and readily understood from 
the kinetic energy gain due to nearest-neighbor hopping processes on all 
four bonds around one hole. Furthermore, applying periodic rather than 
open BCs leads to a division of large corrals into small, diagonal 
ones, which appear to form a lattice (Figs.~4(b), 5(a)). We may conclude 
that the vertical corral shown in Fig.~3(a) is constrained by the system 
dimensions, an observation confirmed at still higher filling. 

\subsection{Variation of temperature}

We have in addition considered the real-space HF approximation at finite 
temperature. This part of the analysis was motivated by the picture of 
the self-consistency procedure as the following of a path through a 
free-energy ``landscape''. Because the solution is a stationary point, 
and not a global minimum, one would wish to sample as many local minima 
as possible, which is facilitated when the landscape be as smooth as 
possible. On descending from high temperatures, one expects a uniform charge 
distribution to be the state of minimum free energy, and that the leading 
instability to an inhomogeneous distribution would appear as a single 
minimum before the landscape could become more complex at lower temperatures. 
The effect of a finite temperature is readily incorporated by a thermal 
factor in each of the mean-field equations (\ref{ehfev}), and by replacing 
the condition on the number of quasiparticle states filled with an equation 
determining the Fermi energy. 

\medskip
\begin{figure}[t!]
\mbox{\psfig{figure=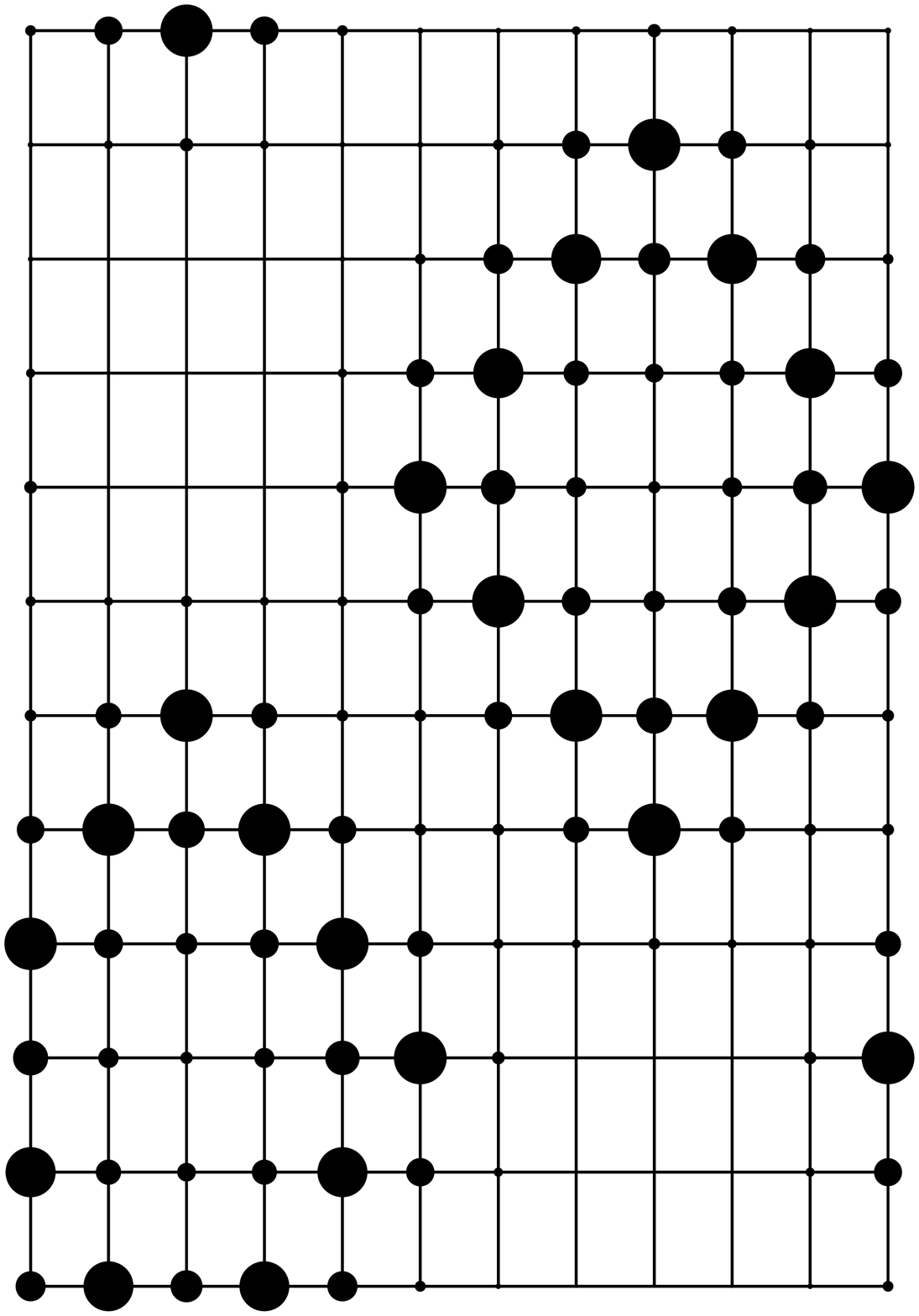,height=5.25cm,angle=0}}
\mbox{\hspace{-0.5cm}\psfig{figure=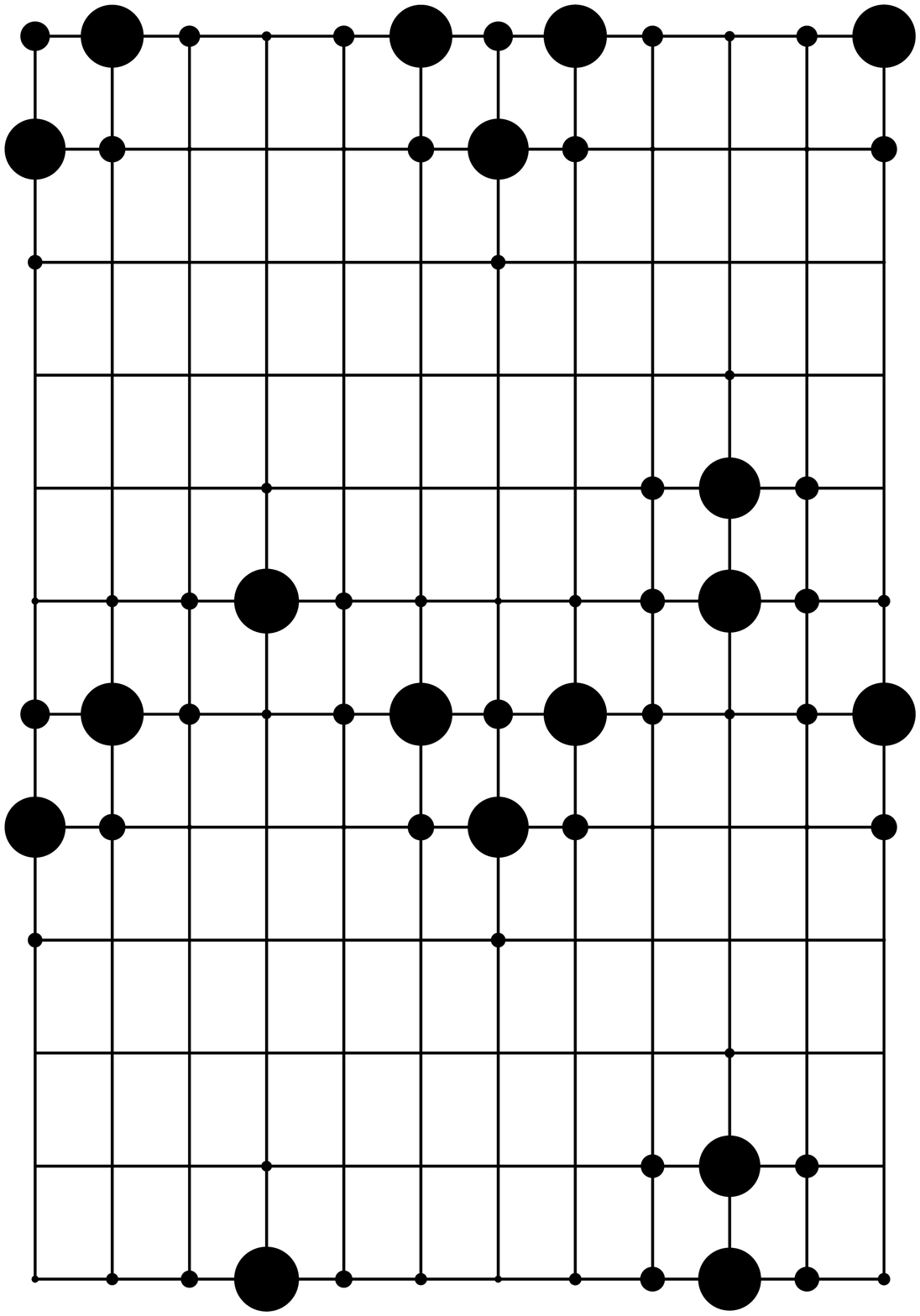,height=5.25cm,angle=0}}
\medskip
{\centerline{(a) \qquad\qquad\qquad\qquad\qquad\qquad (b)}}
\medskip
\caption{Ground-state charge distribution for Hubbard models on a $12 
\times 12$ cluster. (a) $U = 5$, $x = 1/6$, periodic BCs, $t_x = t_y 
= - 1$. (b) $U = 8$, $x = 1/8$, periodic BCs, $t_x = - 1.1$, $t_y = - 0.9$. 
Charge scale as in Fig.~3.  }
\end{figure}

At high temperatures, $T > 2$, the solutions are always uniform in charge 
and have no magnetization. Depending on the exact choice of the other 
parameters, for temperatures in the range $0.18 < T < 2$, an AF spin 
configuration develops while the charge distribution remains uniform. 
Finally, at temperatures $T \sim 0.18$ for most of the situations we 
have considered, an inhomogeneous charge structure develops (corrals, 
polarons, stripes with suitable anisotropy) while the spins remain locally 
AF. In all the cases we considered, the leading instability to charge 
inhomogeneity at finite temperature was identical to the ground-state 
structure found by working entirely at $T = 0$, with a single exception 
mentioned below. This result is reassuring in the sense of helping to 
confirm that the minima emerging from the zero-temperature studies are 
quite likely to be global in nature, and in this work we will not pursue 
the finite-temperature studies beyond establishing this point. We note 
finally that in this approach the onset temperature for formation of an 
incommensurate spin structure where this arises (e.g. Fig.~3(e)) is 
identical to that for charge order. 

\subsection{Variation of anisotropy}

We turn now to the most important aspect of our analysis. Figs.~3 and 
4 (c)-(f) illustrate the effects of increasing the hopping anisotropy 
$\epsilon_t = t_x / t_y - 1$ to 11\% and 22\% for a selection of initial 
parameters and BCs. Figs.~3(a), (c) and (e) give the clearest demonstration 
of stripe formation as a result of hopping anisotropy; although this is for 
open BCs the results for periodic BCs are almost identical for this choice 
of $U$ and $x$. Figs.~3(b), (d) and (f) show that for stronger $U$ the 
polaronic state is more favorable, and although it becomes progressively 
more elongated with increasing anisotropy, stripe formation is not achieved 
at 22\%. Similarly, Figs.~4(a), (c) and (e) show the tendency of corrals to 
become increasingly diamond-shaped with anisotropy, before crossing to a 
fully 1d state. The difference between these and the analogous cases in 
Fig.~3 is that for the smaller filling the (diagonal) corrals are rather 
more stable, and a greater anisotropy would be required to create the 
stripe state. 

One of our key qualitative results is evident in these figures. The 
direction of the stripes, and the major axis of the diamonds, is 
perpendicular to the direction of the strong hopping. Stripes are 
stabilized not by charge motion along them (the picture of a conducting 
channel), but by transverse hopping,\cite{rzo,rcn} onto and off the 
stripe. This is readily shown by a simple argument based on site charge 
densities, which we illustrate for the 4-site cluster with two different 
densities $n_1$ and $n_2$ of Fig.~6. For both spin species ($n_{\uparrow}, 
n_{\downarrow} \equiv n$), a hopping amplitude proportional to the 
occupation of the initial state and the availability of the empty state 
gives 
\begin{mathletters}
\label{esha}
\begin{eqnarray}
\langle E_x^t \rangle / t& = & n_1 (1 - n_1) + n_2 (1 - n_2)  
\nonumber \\ & = & 2 ( n_1 + n_2 - (n_1^2 + n_2^2) ) \\
\langle E_y^t \rangle / t & = & n_1 (1 - n_2) + n_2 (1 - n_1) \nonumber 
\\ & = & 2 (n_1 + n_2 - 2 n_1 n_2 ) , 
\end{eqnarray}
\end{mathletters}
whence the identity $n_1^2 + n_2^2 \ge 2 n_1 n_2$ ensures that 
$\langle E_y^t \rangle \ge \langle E_x^t \rangle$. That hopping between 
two different charge densities is favored over that between equal 
densities ensures that transverse charge fluctuations determine the 
orientation of uniform stripe states such as those in Fig.~3(e). 

This result should not be confused with the metallic or insulating nature 
of the stripes: these are still separated in the transverse direction 
by insulating, AF regions. Of all the charge-inhomogeneous solutions 
illustrated here, only the stripe solutions are metallic, and this in 
the stripe direction within the 2d structure. While we have considered 
transverse hopping as uncorrelated, virtual processes, other microscopic 
approaches suggest a coherent hopping of stripe segments of lengths 
around 5 lattice sites.\cite{rlglv1}

\medskip
\begin{figure}[t!]
\centerline{\psfig{figure=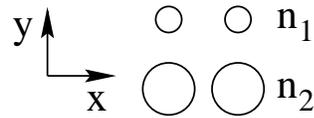,height=1.5cm,angle=0}}
\medskip
\caption{ Four-site cluster with two separate charge densities $n_1$ and 
$n_2$, to illustrate kinetic energy difference between ${\hat x}$- and 
${\hat y}$-directions (Eq.~(\protect{\ref{esha}})). }
\end{figure}

The above reasoning clarifies the somewhat unexpected result in Figs.~4(d), 
4(f) and 5(b): here the stripes are aligned in the direction of the 
stronger hopping. In this case the hole count at $x = 1/8$ is insufficient 
for uniform stripe formation on a $12 \times 12$ cluster, and instead we 
find a non-uniform charge configuration. Such alternating stripes were 
presented in Fig.~1(c) of Ref.~\onlinecite{rgro} (to be contrasted with 
the variant in Fig.~1(a) of the same work), and are one 
HF analogue of ``half-filled'' stripes.\cite{rzo} We will follow this 
terminology, but note in passing that the idealized stripe with half of 
its sites occupied by holes is in fact quarter-filled with electrons. The 
fact that hopping along the stripe proceeds between sites of high and low 
charge density helps to explain (\ref{esha}) the orientation of this type 
of 1d charge inhomogeneity. 

An enduring problem for HF studies of the Hubbard model has been that 
these show a definite preference for uniform, filled stripes, as in 
Fig.~3(e) where the hole density per site along the stripe is unity. 
By contrast, the experimental situation favors half-filled stripes, 
in the intermediate doping range.\cite{rtsanu,rtainun} While there is to 
date no information to distinguish between the possibilities of uniform 
or alternating (Fig.~4(b)) hole distribution within such a stripe, one 
may expect the latter to be favored by increasing $U$. Some 
analytical\cite{rgro,rlglv1} and numerical\cite{rwsl} studies have 
addressed the issue of the additional physics which may be responsible 
for this result. Here we have found that hopping anisotropy represents 
an additional factor which may contribute to the stability of non-uniform, 
half-filled stripes. Although this phase is obtained at the ``magic'' 
filling $x = 1/8$, close to which charge-ordering effects in experiment 
are strongest,\cite{riutngls} we hesitate to comment on a connection due 
to the issue of commensuration effects on the finite cluster. 

In common with other HF studies, this state is somewhat more fragile than 
the uniform stripe, as indicated by the fact that it is preferred over a 
corral only in the presence of periodic BCs, and has a smaller energy gain 
per hole than the uniform stripe (Table III below). This case (Fig.~4(f)) 
was the only situation in which the leading instability at high temperatures 
(a corral) was different from the low-$T$ structure. However, our result 
remains indicative of the possibilities allowed in the presence of hopping 
anisotropies, and we believe representative of the situation for appropriate 
fillings. At larger values of $U$ one observes the emergence of the same 
type of state, but also (Fig.~5(b)) that competition is more severe 
from a phase of isolated, small polarons (Fig.~1(d), Ref.~\onlinecite{rgro}). 
Increasing anisotropy promotes alignment of these polarons towards
the non-uniform stripe phase. 

\medskip
\begin{figure}[t!]
\centerline{\psfig{figure=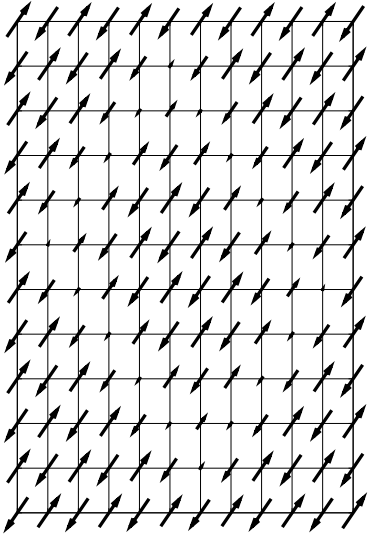,height=7.0cm,angle=0}}
\medskip
\centerline{(a)}
\medskip
\centerline{\psfig{figure=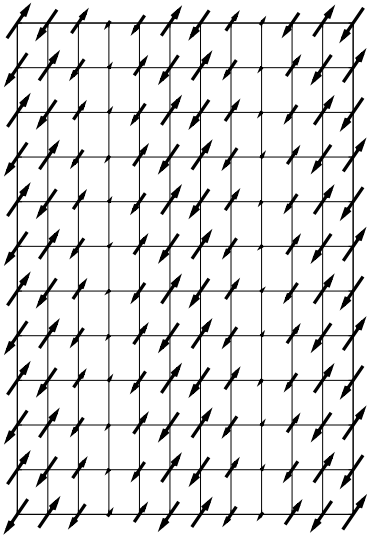,height=7.0cm,angle=0}}
\medskip
\centerline{(b)}
\medskip
\caption{Ground-state spin configuration for Hubbard models on a $12 
\times 12$ cluster with $U = 5$ and open BCs. (a) $x = 1/8$, $t_x = t_y 
= - 1$; corresponding charge distribution in Fig.~4(a). (b) $x = 1/6$, 
$t_x = - 0.9$, $t_y = - 1.1$; corresponding charge distribution in 
Fig.~3(e). Length of largest spins represents 85\% of full moment. }
\end{figure}

A further qualitative point concerns the magnetic nature of the domain 
walls. As illustrated in Fig.~7, all of the corral and uniform stripe 
phases are found to have antiphase domain walls. Those sites located in 
the walls possess no magnetic moment, and the AF regimes on each side 
have a mutual phase shift of $\pi$.\cite{rzg,rpr} Fig.~7(a) shows the 
spin configuration for the diagonal corral state in Fig.~4(a), 
illustrating from the phase shift between the inside and outside of the 
corral that the diagonal domain wall, whose sites have no moment, can be 
considered to be filled, or uniform. Fig.~7(b) shows that the uniform 
stripe state in Fig.~3(e) possesses the same property. By contrast, we 
see in Fig.~8 for the non-uniform stripe state close to half-filling 
that there is no phase change ({\it cf.} Fig.~1(c), Ref.~\onlinecite{rgro}), 
meaning that the stripe is an in-phase domain wall.\cite{rcn} This result 
is also opposite to that of the uniformly half-filled stripe, which is 
an antiphase domain wall\cite{rgro,rwsl} for the same reason as for the 
filled stripe. In the non-uniform case, the sites in the center of the 
stripe retain a magnetic moment, and these are ferromagnetically (FM) 
aligned. However, we believe this to be driven not by the preference 
for a FM hopping channel, but by the spin orientations of the sites 
surrounding those with higher hole densities: these sites are analogous 
to a missing electron, with charge and spin which reflect the hopping 
from the four nearest neighbors. The one-particle physics suggested by 
these observations supports an interpretation by Nagaoka's 
theorem.\cite{rn,rwa} It appears that the non-uniform stripe could be 
considered as aligned, small FM polarons. 

\medskip
\begin{figure}[t!]
\centerline{\psfig{figure=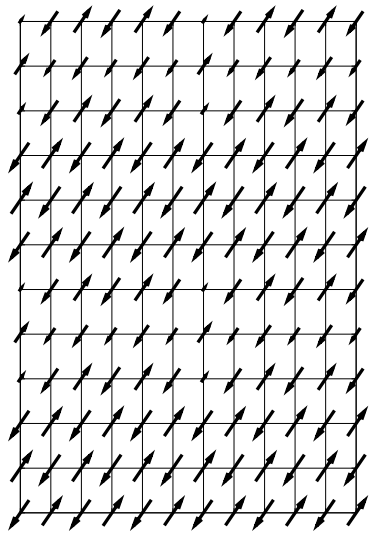,height=7.0cm,angle=0}}
\medskip
\caption{Ground-state spin configuration for Hubbard model on a $12 
\times 12$ cluster with $U = 5$, $x = 1/8$, periodic BCs and $t_x = 
- 0.9$, $t_y = - 1.1$; corresponding charge distribution in Fig.~4(a). 
Spin scale as in Fig.~7. }
\end{figure}

Within the limits of the HF approximation, we may quantify some of 
the above observations by inspecting the energies of the solutions. 
Table I illustrates the effect of increasing anisotropy on the total 
energy for one parameter set (Figs.~3(a), (c) and (e)) by decomposition 
into potential and kinetic energy components. We see that increasing 
anisotropy causes a small increase in the potential contribution 
(double-occupancy energy), but that this is more than compensated by 
gains in the kinetic part. The strong effect of the hopping anisotropy 
is clearly demonstrated by the fact that for a 22\% anisotropy in $t$ 
values (right column), the kinetic energies differ by almost a factor 
of 2. This strong increase in energy difference reflects the dominance 
of transverse hopping in stabilizing the uniform stripe phase. 

\medskip
\begin{table}[t!]
\narrowtext
\caption{Total ground-state energy $E_{\rm tot}$, and kinetic ($E_K^x$, 
$E_K^y$) and potential ($E_P$) energy components per site for Hubbard 
model on a $12 \times 12$ cluster with $U = 5$, $x = 1/6$, open BCs and 
increasing anisotropy $t_x / t_y$; corresponding charge distributions 
in Fig.~3(a), (c) and (e). }
\medskip
\begin{tabular}[hp]{c|c|c|c}
& $t_x / t_y = 1$ & $1.05 / 0.95$ & $1.1 / 0.9$ \\ \hline
$E_{\rm tot}$ $\,$ & $\,$ -0.78782 $\,$ & $\,$ -0.79361 $\,$ & $\,$ -0.80317 \\
$E_K$ $\,$ & $\,$ -1.22931 $\,$ & $\,$ -1.24634 $\,$ & $\,$ -1.25942 \\
$E_K^x$ $\,$ & $\,$ -0.61465 $\,$ & $\,$ -0.72917 $\,$ & $\,$ -0.80513 \\
$E_K^y$ $\,$ & $\,$ -0.61465 $\,$ & $\,$ -0.51717 $\,$ & $\,$ -0.45429 \\
$E_P$ $\,$ & $\,$ 0.44148 $\,$ & $\,$ 0.45273 $\,$ & $\,$ 0.45625
\end{tabular}  
\end{table}

Table II shows the total energies per site for $12 \times 12$ clusters at 
hole dopings $x$ of 0, 1/8 and 1/6, and Table III the differences between 
doped and undoped systems presented as energy per hole. From Table II, 
systems with periodic BCs have significantly lower total energy, but a 
rather similar energy per hole. The same evolution of the total energy 
with anisotropy as in Table I occurs for all parameter sets with open BCs, 
and also at half-filling ($x = 0$), for the same reason as above. However, 
with periodic BCs a small anisotropy appears in most cases to result in a 
higher overall energy, but larger anisotropy values restore the behavior 
observed with open BCs. This result we ascribe to the competition of 
different possible structures within the system before a truly 1d state 
may be established; periodic BCs vastly increase the degeneracy of 
available solutions, sometimes leading to convergence difficulties, and 
one has less confidence of finding the true ground state. Turning to 
Table III, the value of the energy per hole reflects both the mobility 
of the holes and the gain in potential energy due to a lower probability 
of double occupancy. These trends compete as $U$ is increased, and also 
as $x$ is increased, as a consequence of which the values in the table 
vary rather little; this statement applies also for smaller ($x = 1/12$) 
and larger ($x = 1/4$) doping levels (not shown). 

\medskip
\begin{table}[t!]
\narrowtext
\caption{Ground-state energies per site for Hubbard model on a $12 
\times 12$ cluster. O and P denote open and periodic BCs, anisotropies 
quoted as ratio $t_x/t_y$. }
\medskip
\begin{tabular}[hp]{c|c|c|c|c|c}
BCs $\,$ & $\,$ $U$ $\,$ & $\,$ $t_x/t_y$ $\,$ & $x = 0$ & $1/8$ & $1/6$ 
\\ \hline
O $\,$ & $\,$ 5 $\,$ & $\,$ 1.00 $\,$ & $\,$ -0.6321 $\,$ & $\,$ -0.7514 $\,$ 
& $\,$ -0.7878 \\ 
O $\,$ & $\,$ 5 $\,$ & $\,$ 1.11 $\,$ & $\,$ -0.6340 $\,$ & $\,$ -0.7534 $\,$ 
& $\,$ -0.7936 \\
O $\,$ & $\,$ 5 $\,$ & $\,$ 1.22 $\,$ & $\,$ -0.6397 $\,$ & $\,$ -0.7563 $\,$ 
& $\,$ -0.8032 \\
P $\,$ & $\,$ 5 $\,$ & $\,$ 1.00 $\,$ & $\,$ -0.6820 $\,$ & $\,$ -0.7967 $\,$ 
& $\,$ -0.8325 \\
P $\,$ & $\,$ 5 $\,$ & $\,$ 1.11 $\,$ & $\,$ -0.6840 $\,$ & $\,$ -0.7905 $\,$ 
& $\,$ -0.8330 \\
P $\,$ & $\,$ 5 $\,$ & $\,$ 1.22 $\,$ & $\,$ -0.6902 $\,$ & $\,$ -0.7950 $\,$ 
& $\,$ -0.8551 \\
O $\,$ & $\,$ 8 $\,$ & $\,$ 1.00 $\,$ & $\,$ -0.4294 $\,$ & $\,$ -0.5550 $\,$ 
& $\,$ -0.5999 \\
O $\,$ & $\,$ 8 $\,$ & $\,$ 1.11 $\,$ & $\,$ -0.4306 $\,$ & $\,$ -0.5591 $\,$ 
& $\,$ -0.6008 \\
O $\,$ & $\,$ 8 $\,$ & $\,$ 1.22 $\,$ & $\,$ -0.4341 $\,$ & $\,$ -0.5624 $\,$ 
& $\,$ -0.6113 \\
P $\,$ & $\,$ 8 $\,$ & $\,$ 1.00 $\,$ & $\,$ -0.4659 $\,$ & $\,$ -0.5974 $\,$ 
& $\,$ -0.6429 \\
P $\,$ & $\,$ 8 $\,$ & $\,$ 1.11 $\,$ & $\,$ -0.4672 $\,$ & $\,$ -0.5971 $\,$ 
& $\,$ -0.6401 \\
P $\,$ & $\,$ 8 $\,$ & $\,$ 1.22 $\,$ & $\,$ -0.4710 $\,$ & $\,$ -0.6002 $\,$ 
& $\,$ -0.6453 
\end{tabular}  
\end{table}

In concluding this subsection, we have found that hopping anisotropy in a 
Hubbard model can give rise to two qualitatively different types of stripe 
state, whose appearance depends on the intrinsic system parameters, and 
whose nature is influenced by the commensurability of the filling with the 
cluster size. Uniform stripes have the same charge density and no magnetic 
moment on every site, align perpendicular to the direction of strong 
hopping, represent antiphase domain walls between AF regions on either 
side, and are energetically rather stable. Non-uniform stripes have 
alternating charge density and finite magnetic moment on the central sites, 
align in the direction of strong hopping, and act as in-phase domain walls.
Both types of stripe are stabilized by hopping between 
sites of differing charge density, which in the uniform case means 
transverse hole hopping, but in the non-uniform case results in similar 
contributions from both longitudinal and transverse hole motion. 

\medskip
\begin{table}[t!]
\narrowtext
\caption{Ground-state energies per doped hole for Hubbard model on a 
$12 \times 12$ cluster at fillings $x = 1/8$ and $x = 1/6$. O and P denote 
open and periodic BCs, anisotropies quoted as ratio $t_x/t_y$. }
\medskip
\begin{tabular}[hp]{c|c|c|c|c}
BCs $\,$ & $\,$ $U$ $\,$ & $\,$ $t_x/t_y$ $\,$ & $x = 1/8$ & $x = 1/6$ 
\\ \hline
O $\,$ & $\,$ 5 $\,$ & $\,$ 1.00 $\,$ & $\,$ -0.9539 $\,$ & $\,$ -0.9342 \\ 
O $\,$ & $\,$ 5 $\,$ & $\,$ 1.11 $\,$ & $\,$ -0.9553 $\,$ & $\,$ -0.9575 \\ 
O $\,$ & $\,$ 5 $\,$ & $\,$ 1.22 $\,$ & $\,$ -0.9328 $\,$ & $\,$ -0.9809 \\ 
P $\,$ & $\,$ 5 $\,$ & $\,$ 1.00 $\,$ & $\,$ -0.9176 $\,$ & $\,$ -0.9032 \\ 
P $\,$ & $\,$ 5 $\,$ & $\,$ 1.11 $\,$ & $\,$ -0.8502 $\,$ & $\,$ -0.8938 \\ 
P $\,$ & $\,$ 5 $\,$ & $\,$ 1.22 $\,$ & $\,$ -0.8385 $\,$ & $\,$ -0.9890 \\ 
O $\,$ & $\,$ 8 $\,$ & $\,$ 1.00 $\,$ & $\,$ -1.0052 $\,$ & $\,$ -1.0228 \\ 
O $\,$ & $\,$ 8 $\,$ & $\,$ 1.11 $\,$ & $\,$ -1.0284 $\,$ & $\,$ -0.9909 \\ 
O $\,$ & $\,$ 8 $\,$ & $\,$ 1.22 $\,$ & $\,$ -1.0268 $\,$ & $\,$ -1.0634 \\ 
P $\,$ & $\,$ 8 $\,$ & $\,$ 1.00 $\,$ & $\,$ -1.0523 $\,$ & $\,$ -1.0619 \\ 
P $\,$ & $\,$ 8 $\,$ & $\,$ 1.11 $\,$ & $\,$ -1.0391 $\,$ & $\,$ -1.0376 \\ 
P $\,$ & $\,$ 8 $\,$ & $\,$ 1.22 $\,$ & $\,$ -1.0336 $\,$ & $\,$ -1.0461 
\end{tabular}  
\end{table}

\subsection{Variation of extrinsic parameters}

In the preceeding subsections we have already noted some of the effects 
of the extrinsic parameters, a term we use to refer to cluster dimensions, 
BCs, and commensuration between filling and cluster size. Open BCs cause 
a definite tendency for holes to avoid the edges, where their hopping 
energy is lowered. While for small doping this merely sets the inhomogeneous 
charge structure in the center of the cluster, at larger dopings it can act 
to influence the alignment of domain walls (Fig.~3(a)), particularly at 
smaller values of $U$ where the charge structures are extended. 
Periodic BCs relax this constraint, and thus facilitate the formation of 
domain lines with the optimal orientation (diagonal (Fig.~5(a)) or vertical 
(Fig.~4(d), (f), also as in Fig.~3(e))) for the anisotropy, and clusters of 
the optimal size and configuration (Figs.~4(b), 5(a)). These effects are also 
inseparable from the issue of the commensuration between the cluster size 
and the number of doped holes, which is instrumental in determining the 
type of the polaron lattice (Figs.~4(b), 5(a)), and the filling and nature 
of the stripes (Figs.~3(e), 4(f)), in the cases where these form. As in 
all such studies on finite clusters, the possibility arises of large, 
finite-size gaps in the quasiparticle spectrum (for example, between 
fillings of 24 and 25 holes in the $12 \times 12$ system). However, we 
do not expect that such effects would particularly favor one class of 
charge-inhomogeneous solution over another. 

\medskip
\begin{figure}[t!]
\mbox{\psfig{figure=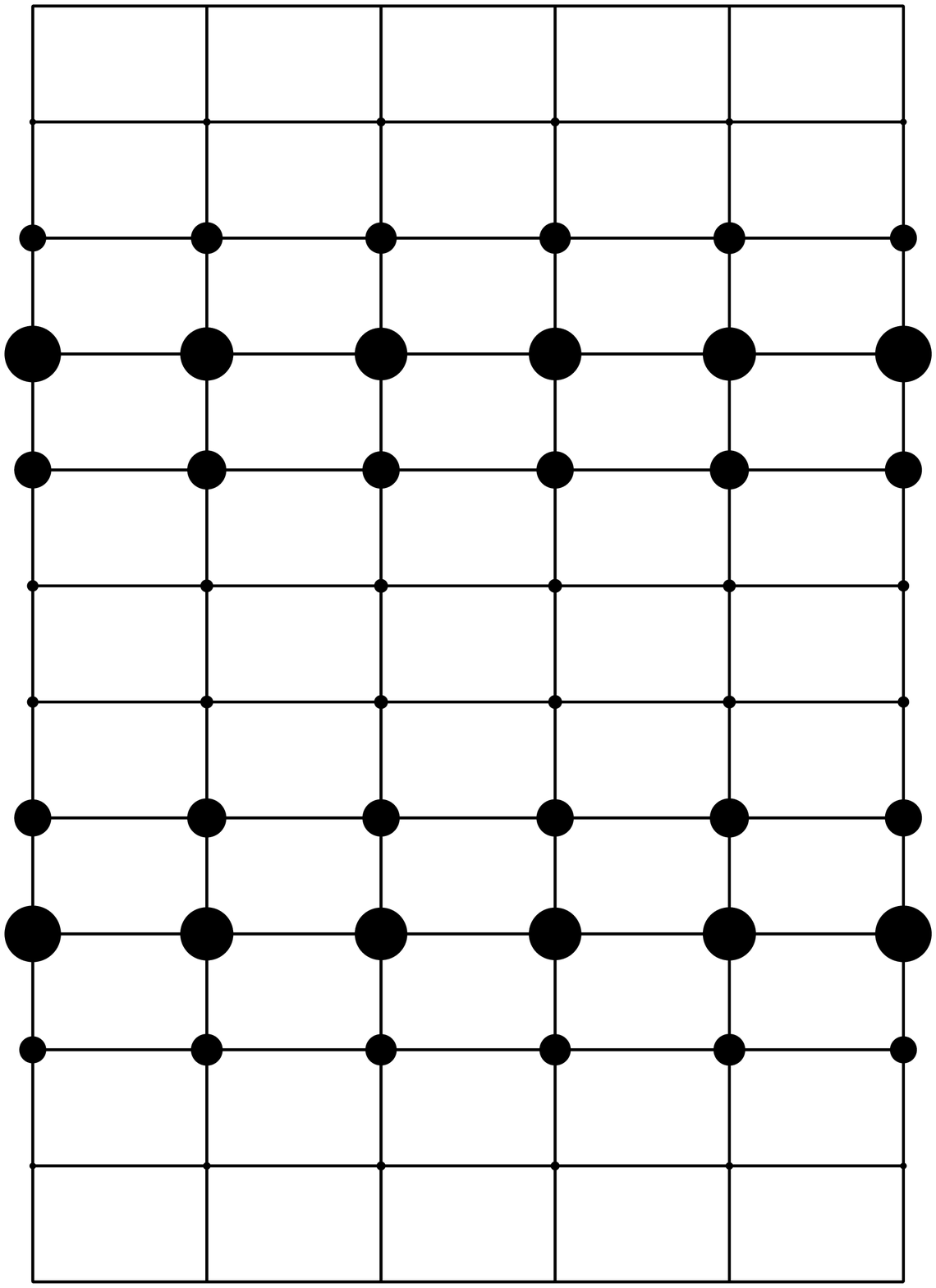,height=5.25cm,angle=0}}
\mbox{\hspace{-0.5cm}\psfig{figure=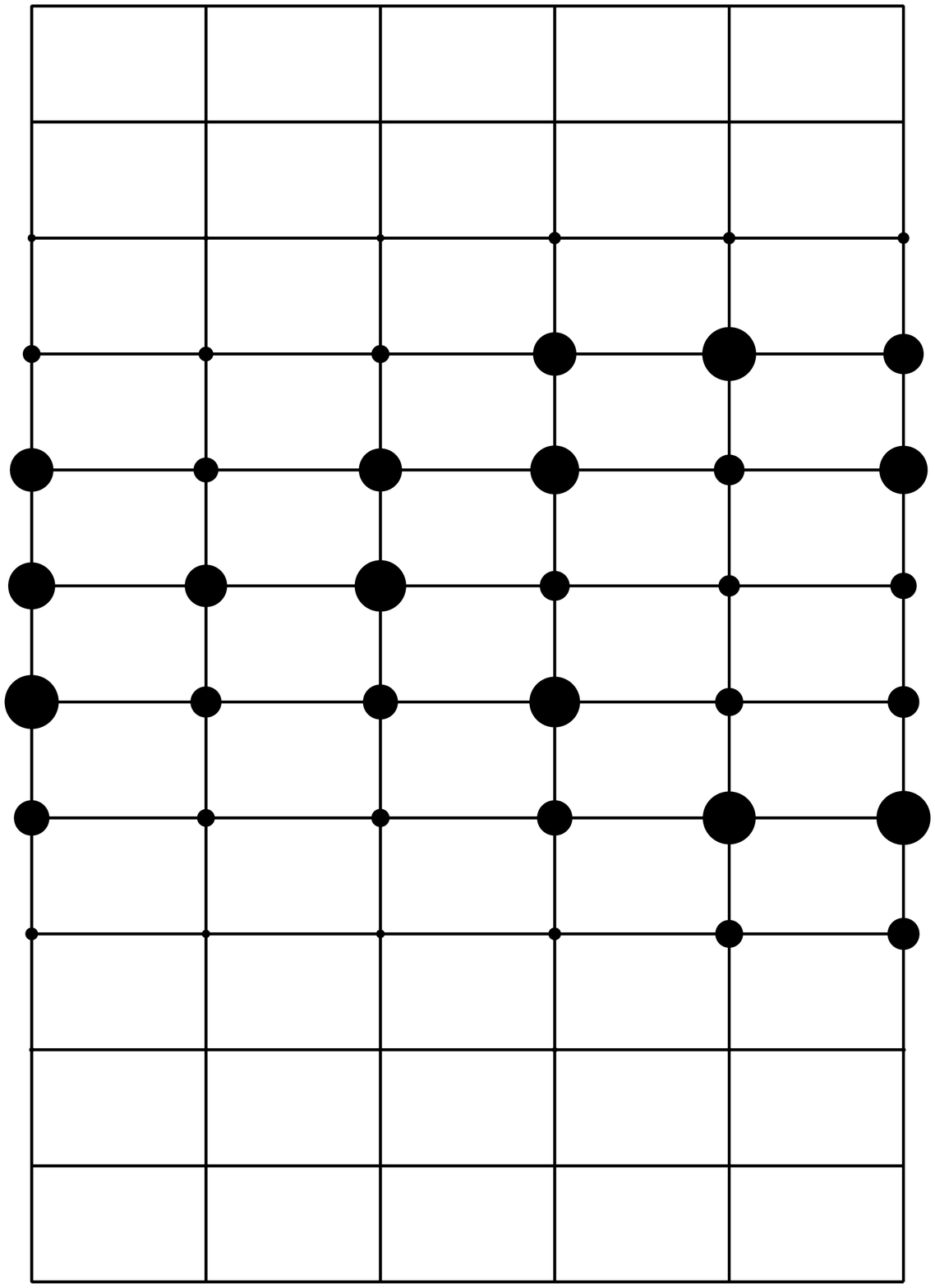,height=5.25cm,angle=0}}
\medskip
{\centerline{(a) \qquad\qquad\qquad\qquad\qquad\qquad (b)}}
\medskip
\caption{Ground-state charge distribution for Hubbard models on a $12 
\times 6$ cluster with $U = 5$ and $t_x = t_y = - 1$. (a) $x = 1/6$, 
open BCs. (b) $x = 1/8$, periodic BCs. Charge scale as in Fig.~3. }
\end{figure}

A more primitive effect of cluster size emerges when at least one of the 
dimensions becomes small on the scale of the equilibrium, inhomogeneous 
charge structure. This is illustrated in Fig.~9(a) for a $12 \times 6$ 
system with hole doping $x = 1/6$. Although the intrinsic parameters are 
the same as in Fig.~3(a), the inhomogeneous charge structure of minimum 
energy in this system is a pair of antiphase domain walls, even for open 
BCs. This observation should be borne in mind when interpreting the 
results of small-system studies such as exact diagonalization and DMRG, 
in which rather more sophisticated analyses are necessary to establish 
the true ground state of a hypothetical infinite system. Fig.~9(b) shows 
the analogous result for $x = 1/8$, where the 9 holes form one and a half 
stripes with periodic BCs. 

Finally, the nature of the ground state is very strongly influenced by the 
choice of periodic or antiperiodic BCs. This is more readily implemented 
by retaining periodic BCs, but working on a cluster with one side chosen 
to be odd, and was the method by which Zaanen and Gunnarsson\cite{rzg} 
first observed stripe solutions in the form of a single domain wall. The 
analogous result for systems even on both sides was a closed domain wall 
like the corrals presented above, but the straight domain line was found 
to offer the minimum energy per particle. We have also considered 
even$\times$odd and odd$\times$odd systems with both periodic and mixed 
BCs (open in the odd direction, periodic in the even\cite{rwsl}), and in 
the same way find the possibility of ``trapping'' stripe solutions in the 
form of single domain walls. Such studies are useful in the context of 
investigating the effects of anisotropy, as one may monitor the total 
energy of the fixed configuration as a function of $t_x / t_y$, and we 
will employ this approach below for the $t$-$J$ model. However, for a 
variety of reasons we do not consider single, infinite domain walls to be 
viable solutions for the isotropic system in the thermodynamic limit, and 
so have chosen not to dwell upon this type of analysis.

For any 3d system of weakly coupled planes, long-range AF order is 
favored if the domain walls form closed loops.\cite{rwsl} We note in our 
studies a preference for domain lines in corrals to lie diagonally, in 
accordance with expectation for nearest-neighbor hopping, but that in no 
cases did a diagonal stripe phase appear with periodic BCs. Instead we 
find a preference for small, diagonal polarons rather than extended lines
in the isotropic system, and that vertical stripes required the LTT-inspired 
hopping anisotropy. A further qualitative observation in this direction 
is the absence in our isotropic, even$\times$even studies of a phase with 
two parallel, vertical domain walls, even at the optimal filling for this 
state ({\it cf.} Fig.~3(e)). With the goal of elucidating the effects of 
anisotropy on the static solutions, we have concentrated on the most 
unbiased initial conditions which appear to retain the possibility of 
sampling the candidate ground states.

\section{\mbox{\bf {\lowercase{$t$}}-$J$} Model}

We turn now to anisotropy effects in the $t$-$J$ model (\ref{ehtj}). The 
previous section has given considerable insight into the complex range 
of possibilities which may arise. Numerical studies of the $t$-$J$ 
model\cite{rwsl} have provided evidence for charge structures of closed 
domain-wall loops similar to those we find in the Hubbard model, but also 
for a close competition with many other structures, including open lines 
(stripes). In order to assess anisotropy effects, and to address the question 
of collaboration or competition between $t$ and $J$ in stripe formation and 
alignment, it will be necessary to characterize the favorability in 
different regimes of 2d {\it vs.} 1d structures, corrals {\it vs.} polarons, 
in-phase {\it vs.} antiphase domain lines, and the possible role of phase 
separation.\cite{rplo,rdr,rpz}

In the real-space HF approximation to the $t$-$J$ model one encounters 
an immediate technical problem. The value of $U$ shold be very large 
(Sec. III), in order to have the Hubbard term act to project out any 
states with doubly-occupied sites from the space of those considered. 
In a selection of studies with $U = 100$, we found the results to vary 
appreciably with changes in $J$, and particularly with changes of $J_x$ 
relative to $J_y$, on the order of 0.05. The ``residual'' superexchange 
term in a pure Hubbard model, $J = 4 t^2 / U = 0.04$ for this choice of 
$U$, suggests that still larger values of $U$ are required for a degree 
of projection which can safely be taken to represent the asymptotic limit. 
This observation is supported by computing the value of the 
double-occupation energy. Indeed, as shown in Fig.~10, while a Hubbard 
model with $U = 100$ gives a robust, polaronic state with variations of 
charge between sites, the situation for $U = 1000$ (and larger) 
corresponds to a cluster of uniformly charged sites more reminiscent of 
incipient phase separation, and with no obvious contribution from kinetic 
terms. We note also in connection with Fig.~10(b) that the calculation 
has converged to a state lacking the fourfold symmetry expected from a 
uniform initial configuration, which is indicative of difficulties in 
employing the real-space HF technique in this parameter regime. 

\medskip
\begin{figure}[t!]
\mbox{\psfig{figure=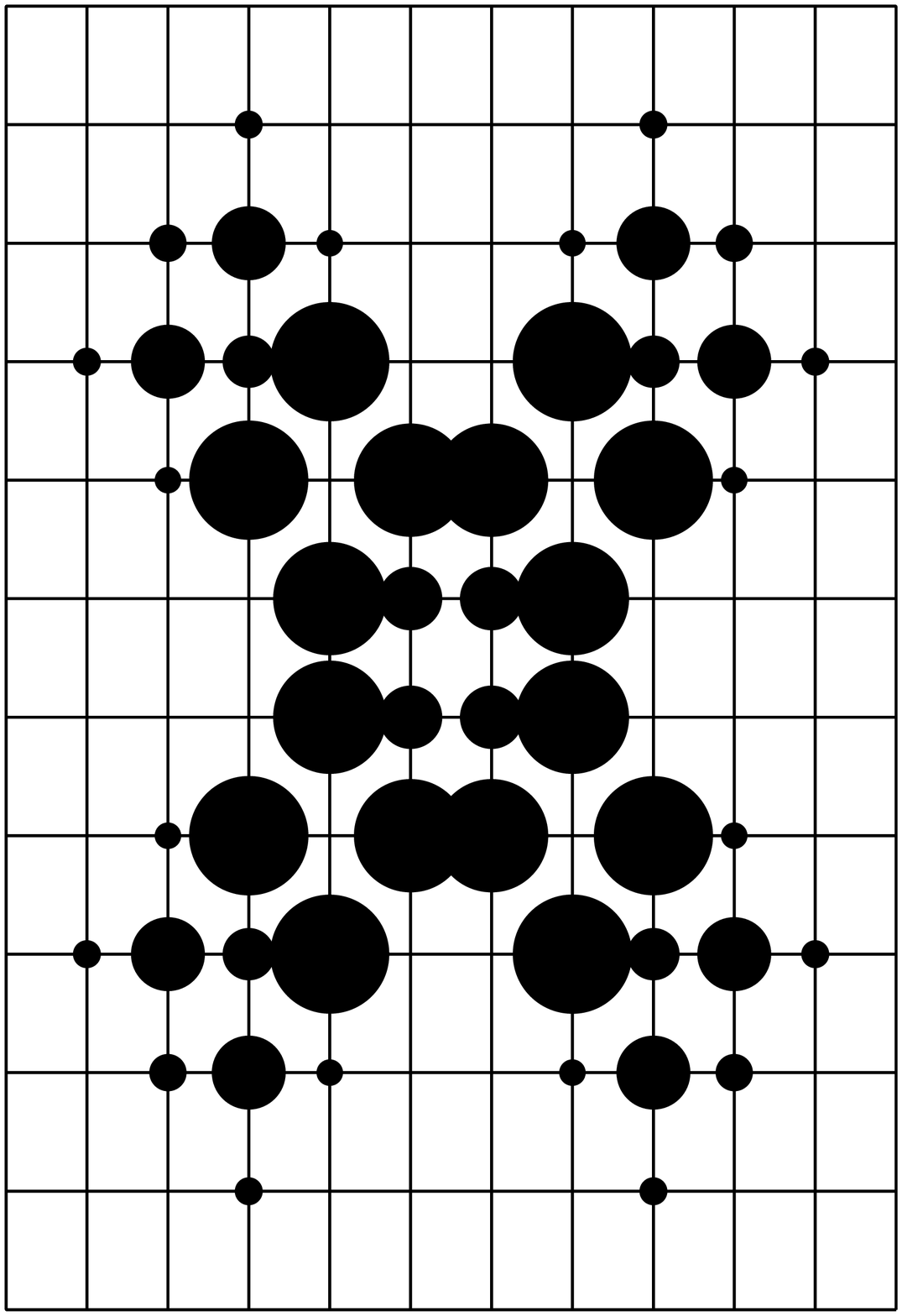,height=5.25cm,angle=0}}
\mbox{\hspace{-0.5cm}\psfig{figure=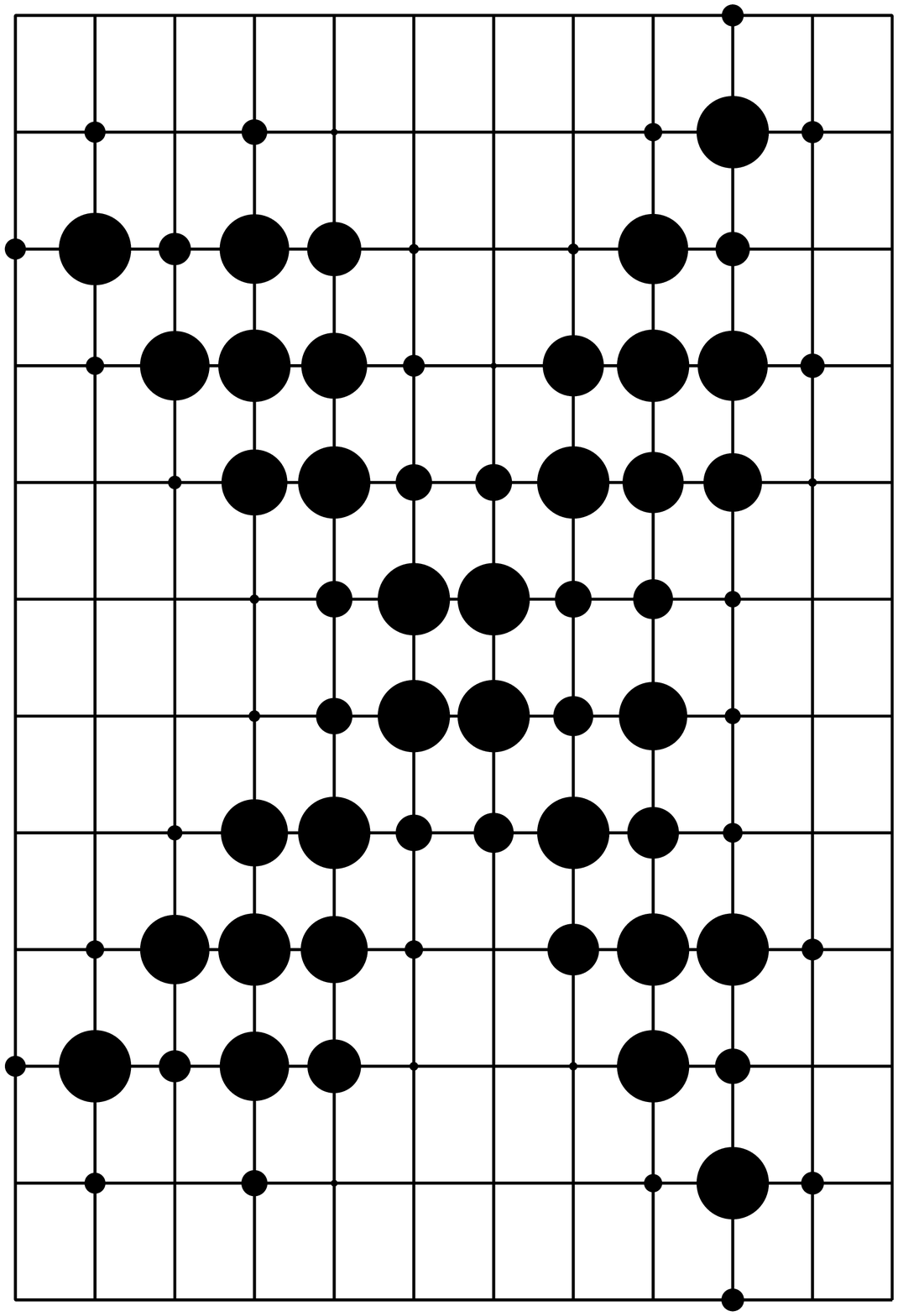,height=5.25cm,angle=0}}
\medskip
{\centerline{(a) \qquad\qquad\qquad\qquad\qquad\qquad (b)}}
\medskip
\caption{Ground-state charge distribution for Hubbard models on a $12 
\times 12$ cluster with $x = 1/6$, open BCs, and $t_x = t_y = - 1$. 
(a) $U = 100$. (b) $U = 1000$. Charge scale as in Fig.~3. Spin 
distributions (not shown) are AF with unaltered phase. }
\end{figure}

For such large values of $U$, (i) every local minimum becomes very deep 
(${\cal O}(U)$), and (ii) the fluctuation energies (${\cal O} (t^2/U)$) 
become very small. Calculations in this regime show that the method loses 
its flexibility to find a global minimum, and that the final state is 
determined largely by the chosen starting configuration. Physically, large 
$U$ projects out the relevance of quantum fluctuations, a statement we will 
make more specific in the following example. There remain two possible 
courses of action by which one may gain insight into the anisotropic $t$-$J$ 
model using the real-space HF approach. As outlined in the preceeding section, 
the first is to create a captive stripe at large $U$, and to study through 
the ground-state energy $E_{\rm tot}$ the qualitative question of how 
anisotropies in $t$ and $J$ affect such a state. The second is to reduce the 
value of $U$ to a size at which the system appears once again to have the 
ability to explore a number of distinct states, and to seek indications for 
the validity of the results in a model with true projection. 

\subsection{Large \mbox{\bf $U$}}

In this subsection we consider the kinetic and magnetic energies of 
filled and half-filled stripes as a function of the anisotropy. We set 
$U = 1000$ to ensure adequate removal of doubly occupied sites, and 
consider the canonical $t$-$J$ model parameter ratio $J / t = 0.35$. We 
trap stripes of 12 and 6 holes along the center line of a $12 \times 11$ 
and $12 \times 12$ clusters, respectively, with periodic BCs, 
and compute the kinetic and magnetic energies of the system 
with $\pm$11\% anisotropy in both $t$ and $J$. Because the stripe is 
aligned along the ${\hat x}$-direction, automatically breaking any 4-fold 
symmetry, positive and negative anisotropy effects are not necessarily 
the same. We find that the schematic picture of the fully local stripe 
is very nearly realised (to 98\%, Fig.~11(a)) in the uniform case (12 
holes) when the spin configuration is set such that the stripe is an 
antiphase domain wall, as in the previous section. Similarly, the 
half-filled stripe solution (Fig.~11(b)) emerges only when the stripe 
is an in-phase domain wall, which accounts for the different transverse 
cluster dimensions used in each case. Thus we do not show the corresponding 
spin distributions for Fig.~11, which consist simply of full-moment AF 
regions on both sides of the stripe, with (a) and without (b) a phase 
shift across the domain wall. 

\medskip
\begin{figure}[t!]
\mbox{\psfig{figure=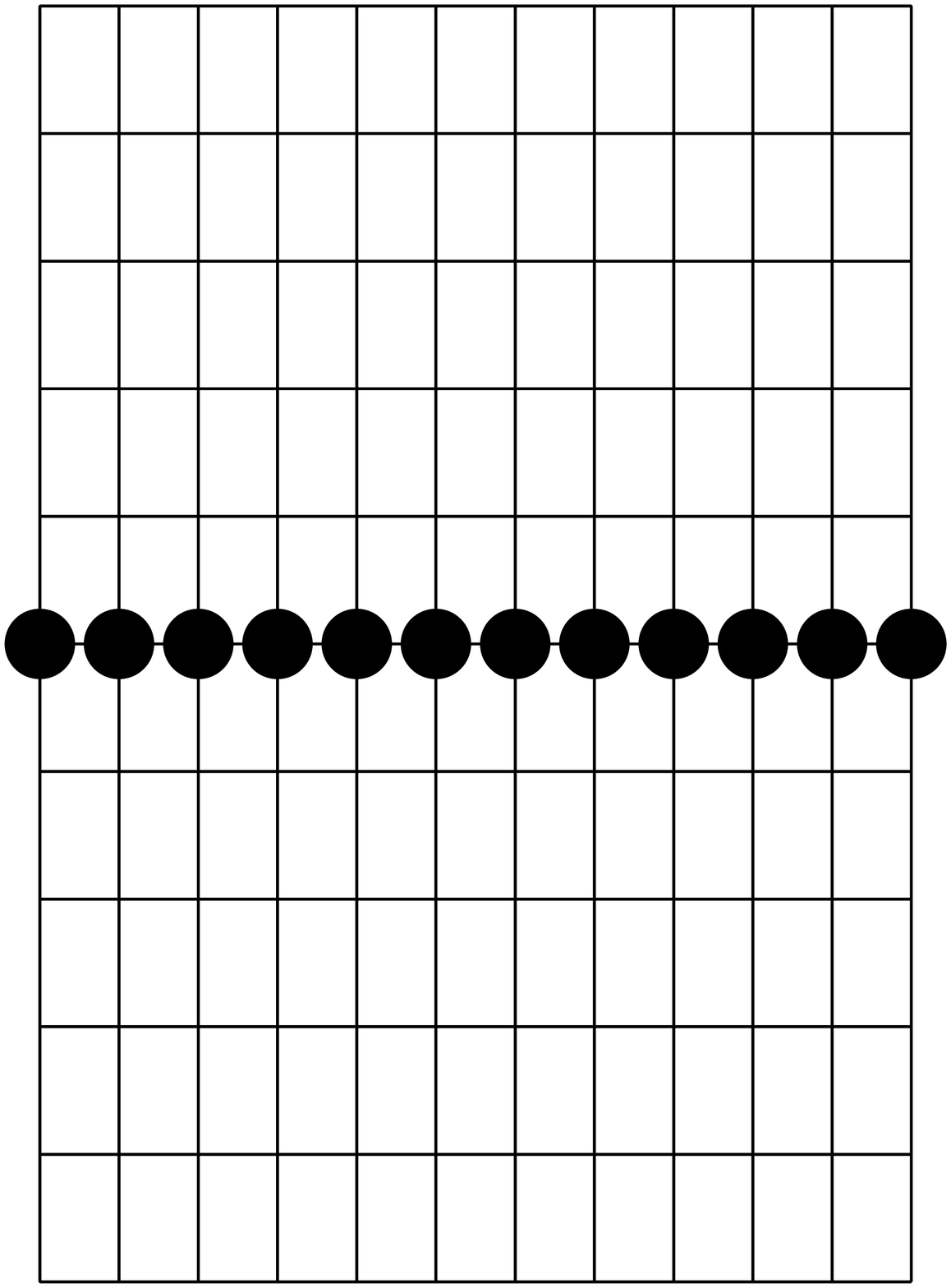,height=5.25cm,angle=0}}
\mbox{\hspace{-0.5cm}\psfig{figure=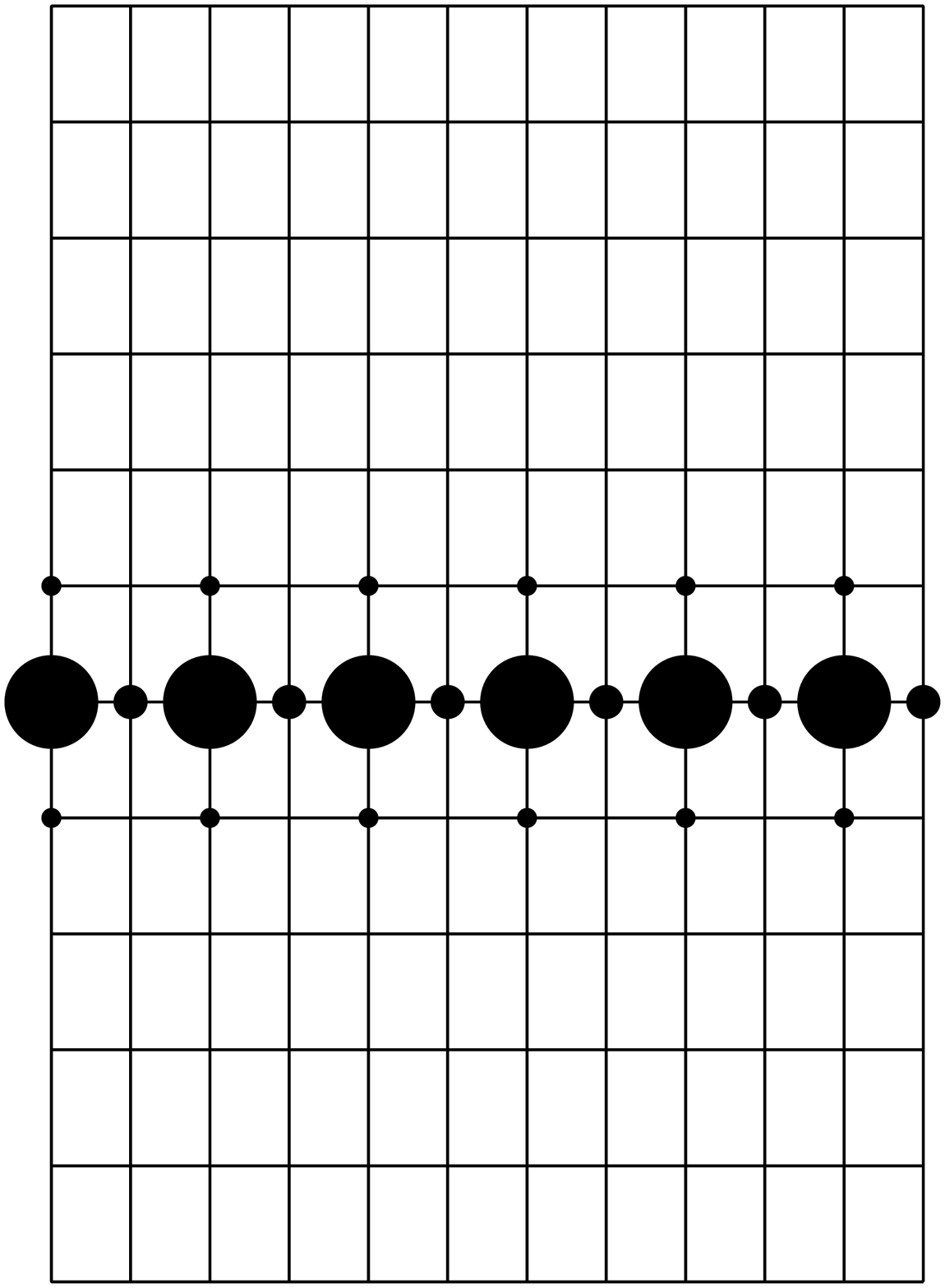,height=5.25cm,angle=0}}
\medskip
{\centerline{(a) \qquad\qquad\qquad\qquad\qquad\qquad (b)}}
\medskip
\caption{Ground-state charge distribution for $t$-$J$ models with 
periodic BCs, $t_x = t_y = - 1$, $J_x = J_y = 0.35$, and $U = 1000$. 
(a) $12 \times 11$ cluster with 12 holes, circle radius on stripe  
represents 98.1\% hole occupation; (b) $12 \times 12$ cluster with 6 
holes, circle radii on stripe represent 55.9\% and 20.3\% hole occupation. }
\end{figure}

The results for kinetic and magnetic energies are presented in Tables 
IV and V. We list these separately because the numbers are not
directly comparable at large $U$: while the magnetic energy is determined 
essentially by $J$, and depends only on the number of antiparallel spin 
pairs, the kinetic energy requires a transfer of electrons between sites. 
Although a transverse hopping would appear to be favorable between 
neighboring filled and empty sites, the expectation value $\langle 
c_{i\sigma}^{\dag} c_{j\sigma} \rangle$ depends on the component of the 
final-state configuration present in the HF wave function (which should 
be considered as a linear superposition of many charge configurations 
in the basis of lattice sites). Large values of $U$ act to suppress 
these additional components, and thus the expectation values $\langle 
E_K \rangle$. This is the sense in which quantum fluctuations 
are projected out, and it is this feature which we will restore in the 
following subsection. 

\medskip
\begin{table}[t!]
\narrowtext
\caption{Kinetic (upper figure) and magnetic (lower) energy components 
per site for $t$-$J$ model on a $12 \times 11$ cluster with 12 holes, 
periodic BCs, and $U = 1000$. }
\medskip
\begin{tabular}[hp]{c|c|c|c}
& $t_x = - 0.95$ & $t_x = - 1 $ & $t_x = - 1.05$ \\
& $t_y = - 1.05$ & $t_y = - 1 $ & $t_y = - 0.95$ \\ \hline
$J_x = 0.3325$ $\,$ & $\,$ -0.045534 $\,$ & $\,$ -0.043226 $\,$ & $\,$ 
-0.040996 \\
$J_y = 0.3675$ $\,$ & $\,$ -0.30086 $\,$ & $\,$ -0.30088 $\,$ & $\,$ 
-0.30090 \\ \hline
$J_x = 0.35$ $\,$ & $\,$ -0.045512 $\,$ & $\,$ -0.043206 $\,$ & $\,$ 
-0.040976 \\
$J_y = 0.35$ $\,$ & $\,$ -0.30163 $\,$ & $\,$ -0.30165 $\,$ & $\,$ 
-0.30167 \\ \hline
$J_x = 0.3675$ $\,$ & $\,$ -0.045491 $\,$ & $\,$ -0.043185 $\,$ & $\,$ 
-0.040955 \\
$J_y = 0.3325$ $\,$ & $\,$ -0.30239 $\,$ & $\,$ -0.30241 $\,$ & $\,$ 
-0.30243  
\end{tabular} 
\end{table}

\medskip
\begin{table}[t!]
\narrowtext
\caption{Kinetic (upper figure) and magnetic (lower) energy components 
per site for $t$-$J$ model on a $12 \times 11$ cluster with 6 holes, 
periodic BCs, and $U = 1000$. }
\medskip
\begin{tabular}[hp]{c|c|c|c}
& $t_x = - 0.95$ & $t_x = - 1 $ & $t_x = - 1.05$ \\
& $t_y = - 1.05$ & $t_y = - 1 $ & $t_y = - 0.95$ \\ \hline
$J_x = 0.3325$ $\,$ & $\,$ -0.089029 $\,$ & $\,$ -0.088635 $\,$ & $\,$ 
-0.088445 \\
$J_y = 0.3675$ $\,$ & $\,$ -0.31249 $\,$ & $\,$ -0.31261 $\,$ & $\,$ 
-0.31272 \\ \hline
$J_x = 0.35$ $\,$ & $\,$ -0.089018 $\,$ & $\,$ -0.088626 $\,$ & $\,$ 
-0.088438 \\
$J_y = 0.35$ $\,$ & $\,$ -0.31255 $\,$ & $\,$ -0.31267 $\,$ & $\,$ 
-0.31280 \\ \hline
$J_x = 0.3675$ $\,$ & $\,$ -0.089006 $\,$ & $\,$ -0.088615 $\,$ & $\,$ 
-0.088429 \\
$J_y = 0.3325$ $\,$ & $\,$ -0.31260 $\,$ & $\,$ -0.31275 $\,$ & $\,$ 
-0.31289  
\end{tabular}  
\end{table}

Comparing the kinetic energies for the uniform stripe configuration 
(Table IV, Fig.~11(a)), we find little variation with $J$, and a strong 
preference for larger transverse hopping $t_y$. Thus the effect of 
hopping anisotropy is as in the Hubbard model. Quantitatively, that 11\% 
anisotropy leads to a 5\% energy change, in contrast to the 30\% effect 
in the Hubbard model with periodic BCs, can be ascribed to the fact 
that the 1d state is already established in the isotropic case. The 
magnetic energies show a negligible dependence on the hopping, and 
a rather small overall effect which is due to the large AF regions 
undisturbed by the presence of the stripe. However, the variation 
clearly indicates a preference for alignment of the stripe with the 
direction of stronger superexchange, a result readily explained at 
zeroth order: the stripe cuts two bonds in the ${\hat y}$-direction, 
and only one in ${\hat x}$, so clearly costs less magnetic energy in 
longitudinal alignment. Thus for filled stripes in the $t$-$J$ model the 
two anisotropies compete. A detailed analysis by techniques better 
able to account for quantum fluctuation effects\cite{rwsl,rwsb} is 
clearly required to quantify this competition.

Results for the half-filled, in-phase stripe (Fig.~11(b)) are given in 
Table V. The variation in kinetic energies with anisotropy is remarkably 
small in comparison with the uniform stripe, and it would appear fair to 
say that leading-order hopping contributions along the chain are cancelled 
by transverse hopping (primarily to the sites of larger hole density) for 
the ideal configuration established here. These results present no evidence 
to support alignment of the alternating stripe with the direction of strong 
hopping, and in fact show a weak preference for transverse alignment. 
The variation in magnetic energies with 
anisotropy is similarly weak, as expected from the bond-counting argument. 
Corrections arise from the fact that the stripe now possesses a FM center 
line ({\it cf.} Fig.~9), and tend also to favor transverse alignment. 
The orientation of half-filled stripes is thus a delicate issue, and we 
are unable to make any firm statements on the basis of these results. For 
the same reason we have not tried to address further interesting questions 
which may be posed in this context, such as the energetics of corners in 
domain walls, crossings of stripes, or interactions between stripes 
placed in proximity to each other.\cite{rzg} Finally, for both types of 
stripe we note that there is negligible interplay between the individual 
effects of anisotropies in $t$ and $J$, with energy changes proceeding 
linearly and separately in both variables.

\subsection{Intermediate \mbox{\bf $U$}} 

In this subsection we will show results obtained in the spirit of setting 
$U$ as large as possible while retaining the possibility of finding the 
true ground state. This is the approach adopted in Ref.~\onlinecite{rvr}, 
and enables one to restore the competition between kinetic and magnetic 
energies in examining the above effects, but introduces a different set 
of questions concerning the physics of the $t$-$U$-$J$ model.\cite{rdsw}  

We begin by characterizing the effects of $J$ in the isotropic model with 
$U = 50$. Fig.~12 shows the evolution of the ground-state charge 
distribution as $J$ is increased. Even rather small values of the 
superexchange interaction alter the $J = 0$ structure (Fig.~12(a)) to 
one with stronger charge differentiation between and within small 
clusters (Fig.~12(b), also realized for $J = 0.05$). At intermediate 
values of $J$ there is evidence of competition with a phase (not shown) 
of small polarons showing preferential diagonal alignment. As $J$ is 
increased towards the canonical cuprate-model value of 0.3, the small-$J$ 
polaronic phase of Fig.~12(b) exhibits increasing spacing of the charge 
clusters, or preference for undisturbed AF regions (Fig.~12(c)), and shortly 
beyond this we find separated clumps of uniformly distributed charges 
(Fig.~12(d)) in a phase which we regard as representative of true phase 
separation. The quantitative value at which this separation occurs is not 
unexpected from previous estimates.\cite{rplo,rdr,rpz} 

\medskip
\begin{figure}[t!]
\mbox{\psfig{figure=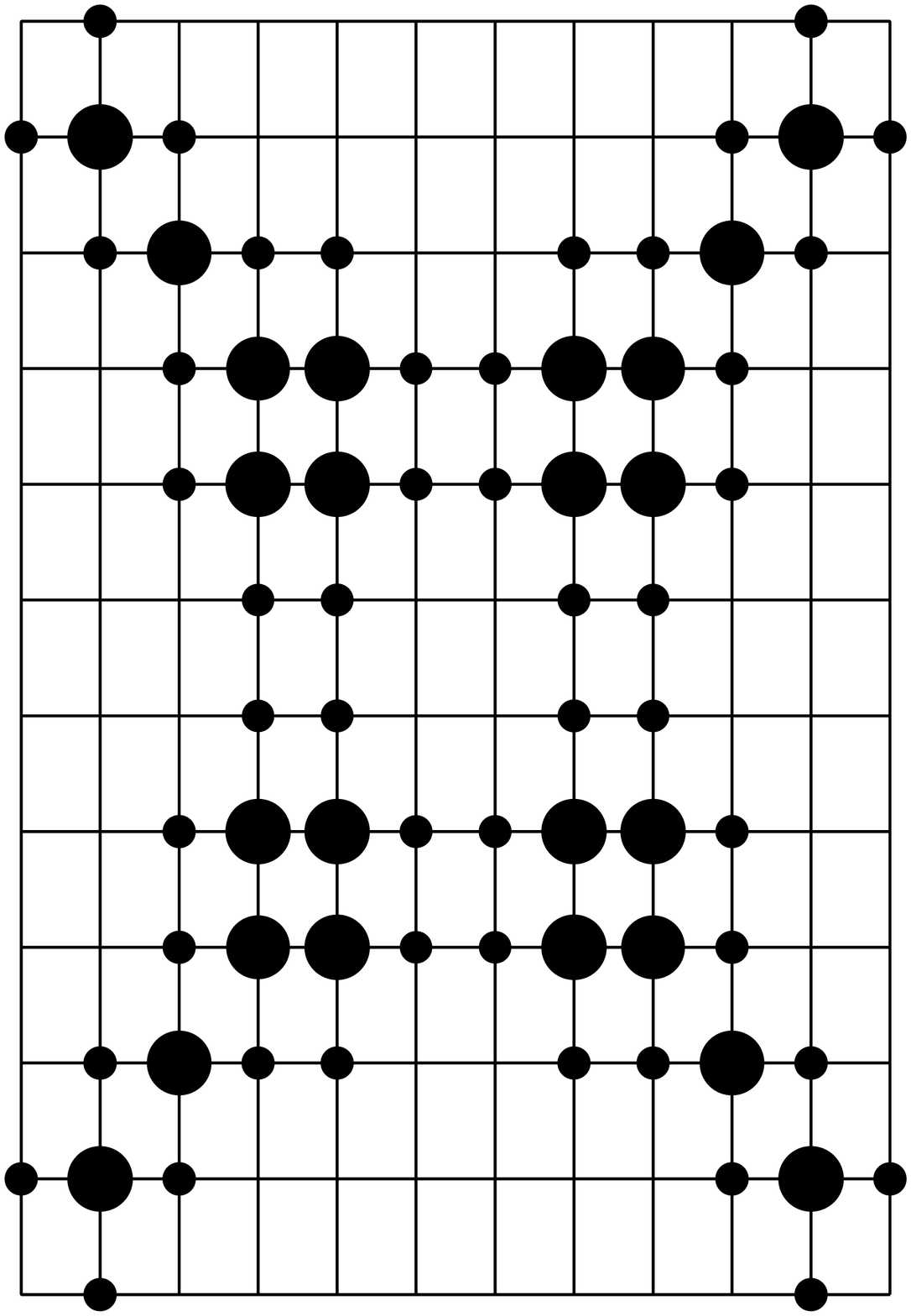,height=5.25cm,angle=0}}
\mbox{\hspace{-0.5cm}\psfig{figure=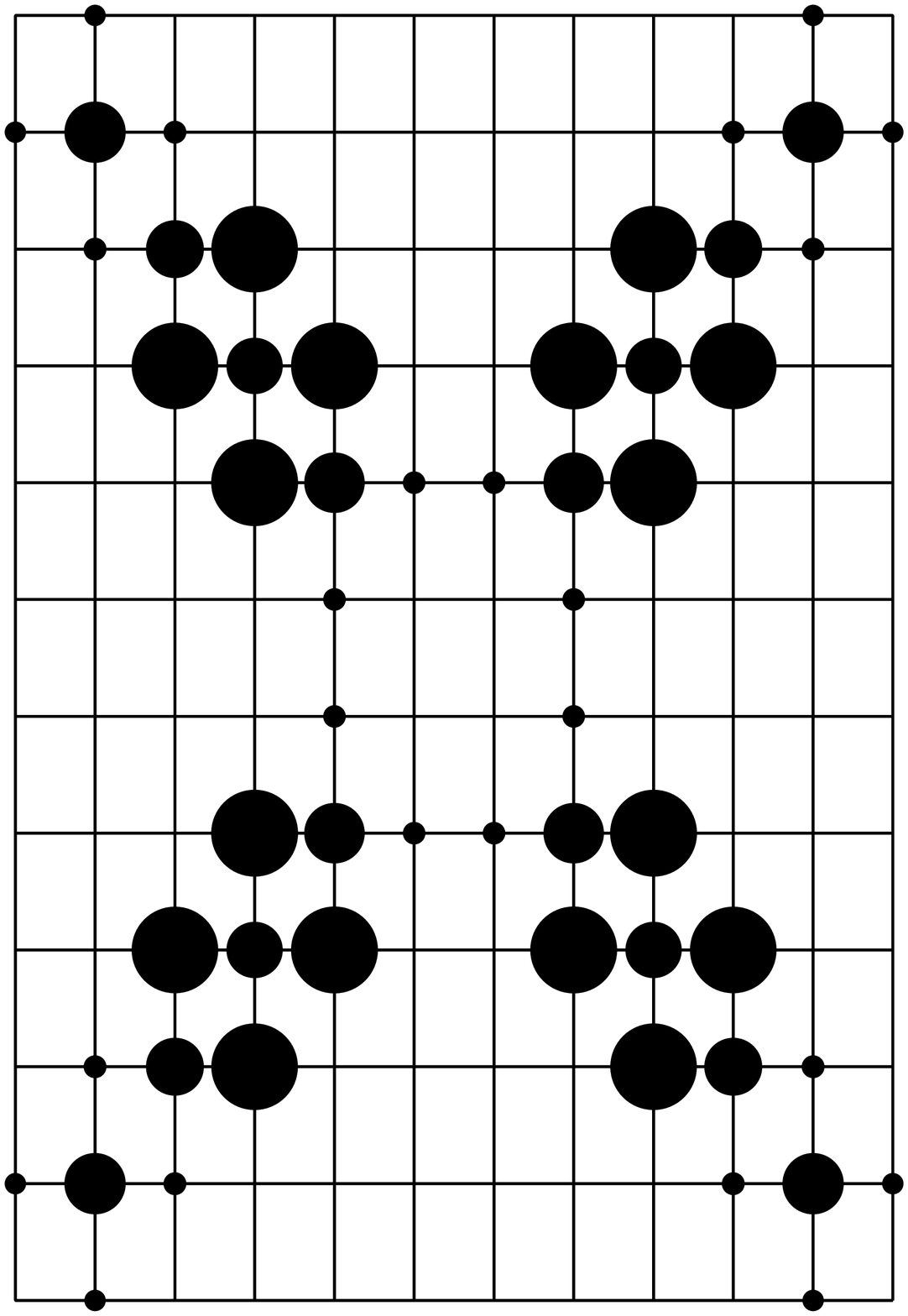,height=5.25cm,angle=0}}
\medskip
{\centerline{(a) \qquad\qquad\qquad\qquad\qquad\qquad (b)}}
\medskip
\mbox{\psfig{figure=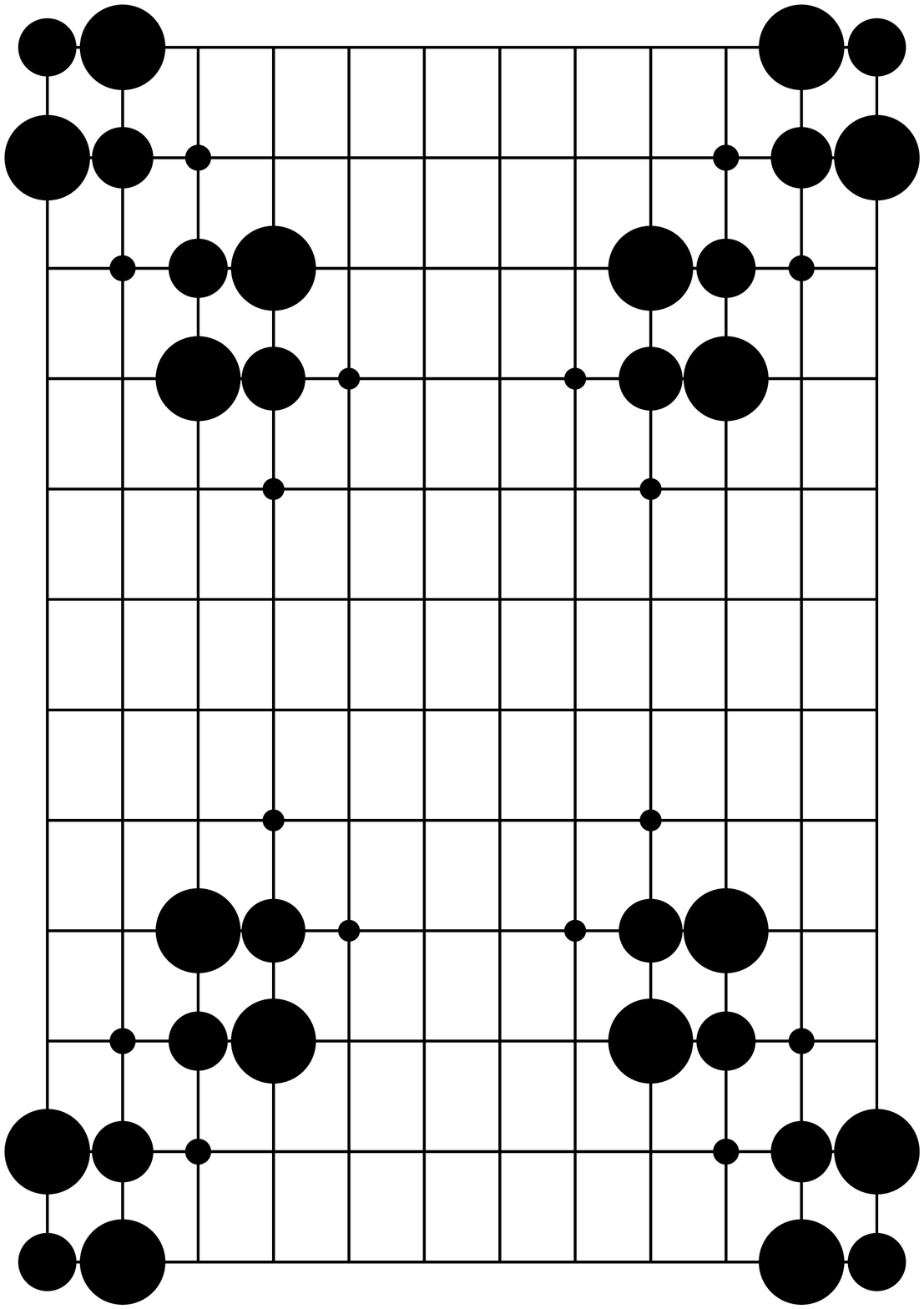,height=5.25cm,angle=0}}
\mbox{\hspace{-0.5cm}\psfig{figure=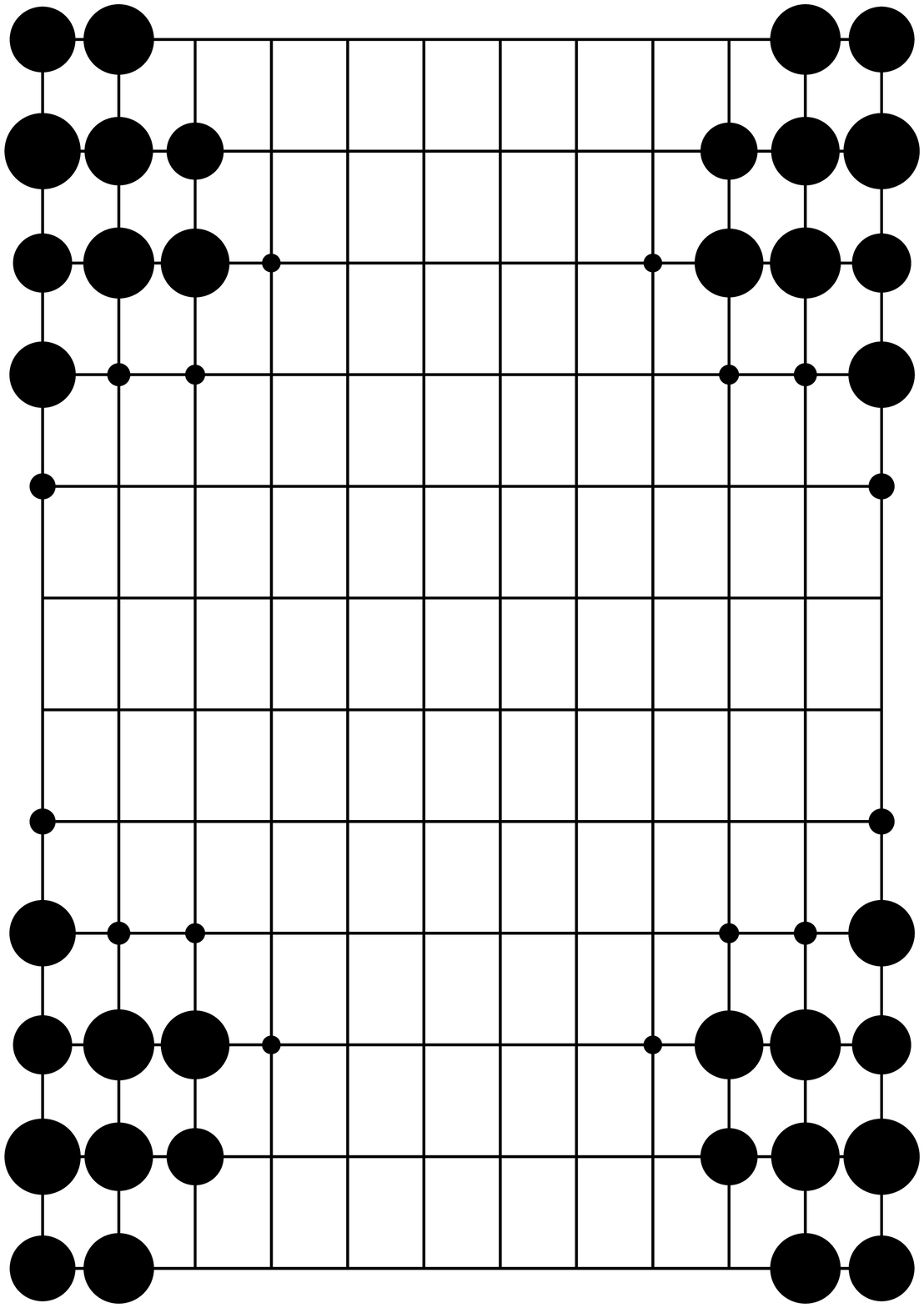,height=5.25cm,angle=0}}
\medskip
{\centerline{(c) \qquad\qquad\qquad\qquad\qquad\qquad (d)}}
\medskip
\caption{Ground-state charge distribution for $t$-$U$-$J$ model on a 
$12 \times 12$ cluster with open BCs, for hole doping $x = 1/6$, $U = 50$, 
and $t_x = t_y = - 1$. (a) $J_x = J_y = 0$. (b) $J_x = J_y = 0.1$. (c) 
$J_x = J_y = 0.3$. (d) $J_x = J_y = 0.5$. Charge scale is such that 
largest circles represent 75.1\% hole occupation. Spin distributions 
(not shown) are AF with unaltered phase. }
\end{figure}

With periodic BCs this tendency to phase separation is strengthened, but 
so also is the tendency to formation of anisotropic structures in the 
presence of hopping anisotropy (below). In certain parameter regimes it 
appears possible for this anisotropy to suppress phase separation, which 
in a systematic study would lead to moving the phase boundary to higher $J$. 
By contrast, anisotropy in the superexchange has little effect at small 
$J (\lesssim 0.1)$, where the system remains largely dominated by kinetic 
processes, but appears to promote the phase separation. 

In this parameter regime the spin state is not altered from its starting 
configuration of very weak antiferromagnetism, which is simply  
reinforced during the self-consistency procedure. This maintenance of the 
spin configuration is not an artefact of considering only a single spin 
component (Sec. III), as retaining all three components returns the 
identical result expected from spin rotation symmetry. While the resulting 
charge inhomogeneities have in-phase nature, we cannot exclude the 
possibility that more extended structures ({\it cf.} Sec. IV), which 
reverse the spins over a finite region, form a separate class of minima 
inaccessible from this initial state. 

\medskip
\begin{figure}[t!]
\mbox{\psfig{figure=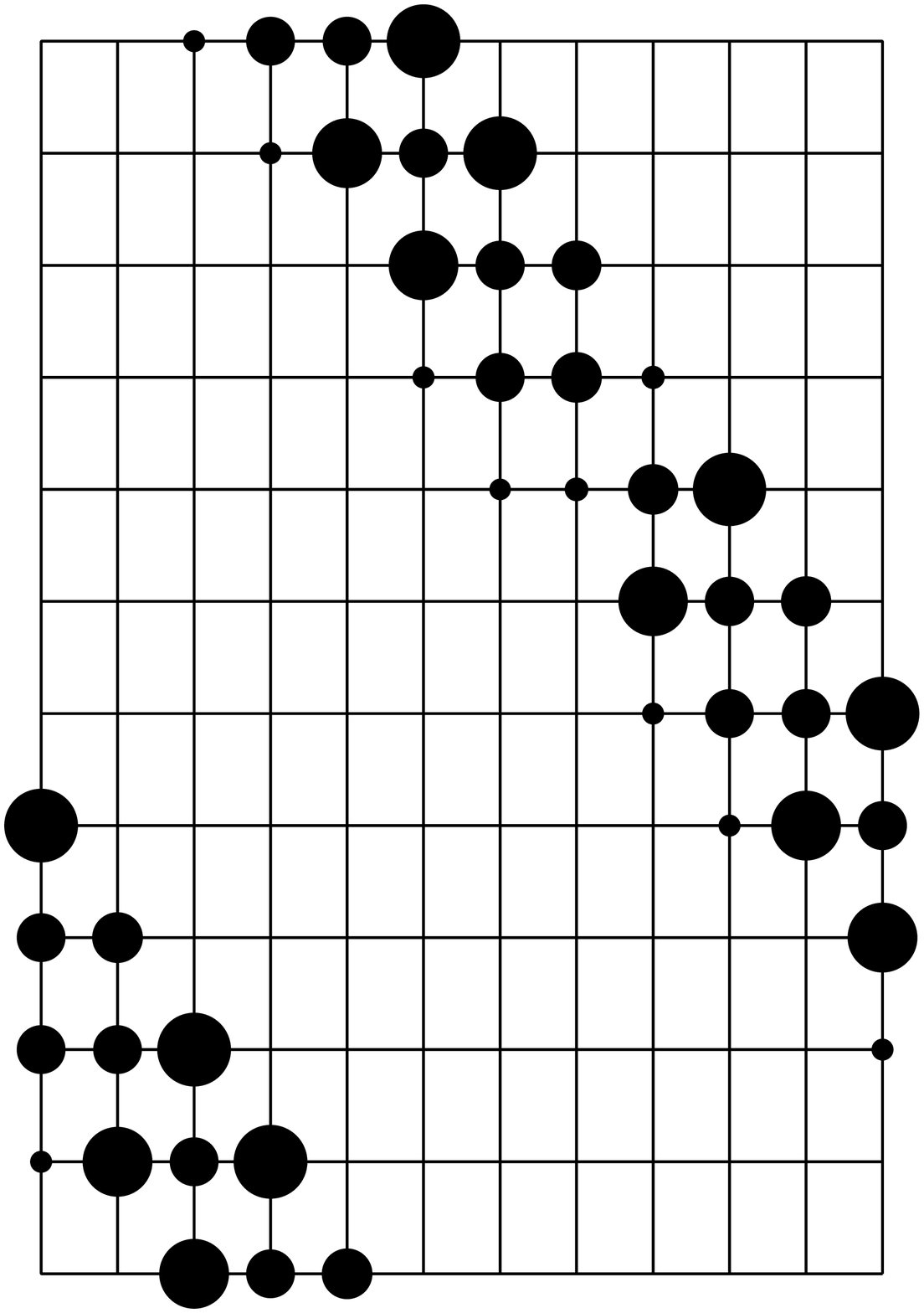,height=5.25cm,angle=0}}
\mbox{\hspace{-0.5cm}\psfig{figure=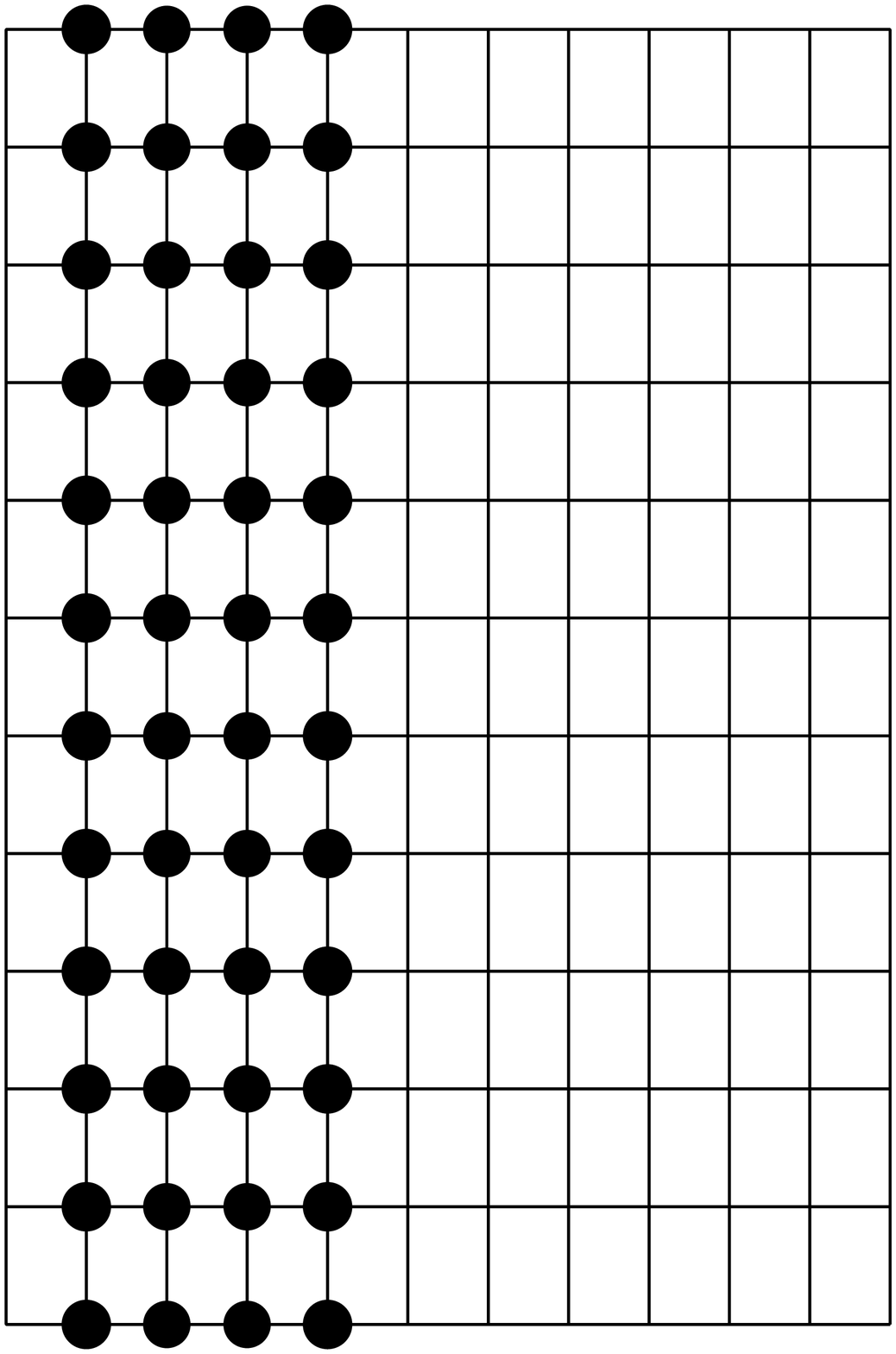,height=5.25cm,angle=0}}
\medskip
{\centerline{(a) \qquad\qquad\qquad\qquad\qquad\qquad (b)}}
\medskip
\caption{Ground-state charge distribution for $t$-$U$-$J$ model on a 
$12 \times 12$ cluster with periodic BCs, for hole doping $x = 1/6$, 
$U = 100$, and $J_x = J_y = 0.2$. (a) $t_x = t_y = - 1$. (b) $t_x = -1.1$,
$t_y = - 0.9$. Charge scale is such that largest circles represent 75.5\% 
hole occupation. }
\end{figure}

Fig.~13(a) illustrates this point with a result obtained 
in an isotropic system with periodic BCs and $U = 100$. The 
emergence of a diagonal stripe is a consequence of the commensurate 
filling, and in fact the structure should more correctly be described 
as a ``bistripe''. The possibility offered by this filling to create a 
wall with a $2 \pi$ phase rotation (Fig.~14) allows this configuration 
to exist within an otherwise in-phase, AF background. Hopping anisotropy 
for this parameter choice leads to a rather different form of bistripe 
(Fig.~13(b)), whose orientation is changed to vertical even by rather 
small differences between $t_x$ and $t_y$. As noted above, superexchange 
anisotropy $J_x \ne J_y$ promotes polaron formation and phase separation 
of the above kind (Fig.~12(f)); quantitatively, 11\% anisotropy at $J = 
0.3$ ({\it cf.} Fig.~12(c)) and 22\% at $J = 0.1$ ({\it cf.} Fig.~12(b)) 
are sufficient to drive the separation. Finally, the qualitative results 
described in this section depend rather little on the chosen value of the 
``large'' $U$, which was varied between 16 and 100. 

\medskip
\begin{figure}[t!]
\centerline{\psfig{figure=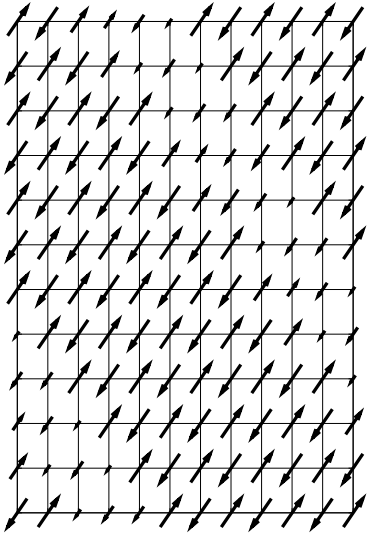,height=7.0cm,angle=0}}
\medskip
\caption{Ground-state spin configuration for $t$-$U$-$J$ model on 
a $12 \times 12$ cluster with periodic BCs, for hole doping $x = 1/6$, 
$U = 100$, $t_x = t_y = - 1$, and $J_x = J_y = 0.2$; corresponding 
charge distribution in Fig.~13(a). Spin scale as in Fig.~7. }
\end{figure}

We return now to the question of the competition between the stripe 
alignment tendencies of anisotropies in $t$ and $J$. The most 
unbiased approach we can find to investigate this is to 
set $U = 16$, and to begin with a uniform charge distribution, and zero 
initial spin on all sites except those in two corners. The following 
results are presented for a $12 \times 11$ cluster with periodic BCs, 
which we have chosen to favor stripes of the experimental orientation. 
We used $J = 0.5$, as in Ref.~\onlinecite{rvr}, but found little 
qualitative difference at $J = 0.25$ despite the results of Fig.~12. 
In Fig.~15(a) is shown the charge distribution for parallel spins at 
diagonally opposite corners of the cluster, which for the chosen 
dimensions sets the requirement of an antiphase domain wall. Indeed a 
domain wall is formed, but despite the commensurate filling (12 holes) 
it is not straight, and wandering of the domain line becomes 
more pronounced at smaller $J$, such that the line sections 
are closer to being diagonal. The competition between a preference for 
diagonal domain walls from nearest-neighbor hopping and vertical domain 
walls from superexchange is exactly that expected from the simplest 
considerations on the idealized stripe. 

In Fig.~15(b) is shown the result of setting an in-phase configuration 
across the cluster: the holes prefer to form a diagonal domain line, and a 
wandering one whose local direction is again mostly diagonal. On periodic 
continuation of the cluster, the line appears as a sawtooth. The prevalence
of kinetic terms over magnetic in the charge alignment is not apparent 
on the basis of Fig.~12, but we note that the doping here is much lower. 
In this case there is a reversal of the AF spin configuration 
across the domain wall, a result which argues for initial configurations 
containing a single ``seed'' spin, from which all other orientations may 
be ``grown''. In fact such an approach produces few qualitative differences 
from the results presented in this section, where the starting configuration 
was weakly AF. For smaller values of $U$ the 
system is sufficiently flexible to choose the appropriate configuration, and 
for larger $U$ the result of the ``growth'' procedure is either a rather 
more diffuse version of the same charge structure (with higher net energy), 
or a random arrangement indicating the local-minimum problem. 

\medskip
\begin{figure}[t!]
\mbox{\psfig{figure=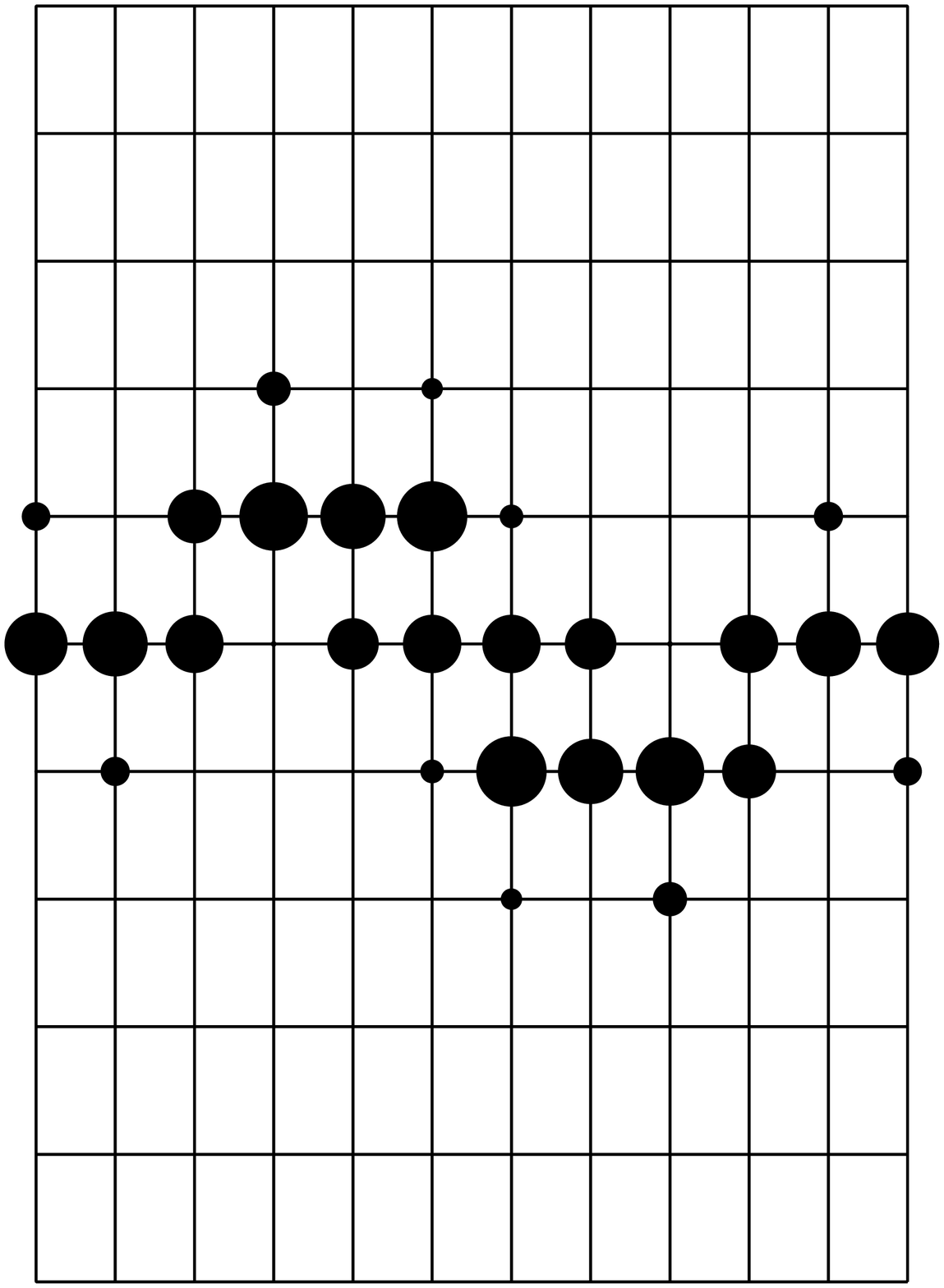,height=5.25cm,angle=0}}
\mbox{\hspace{-0.5cm}\psfig{figure=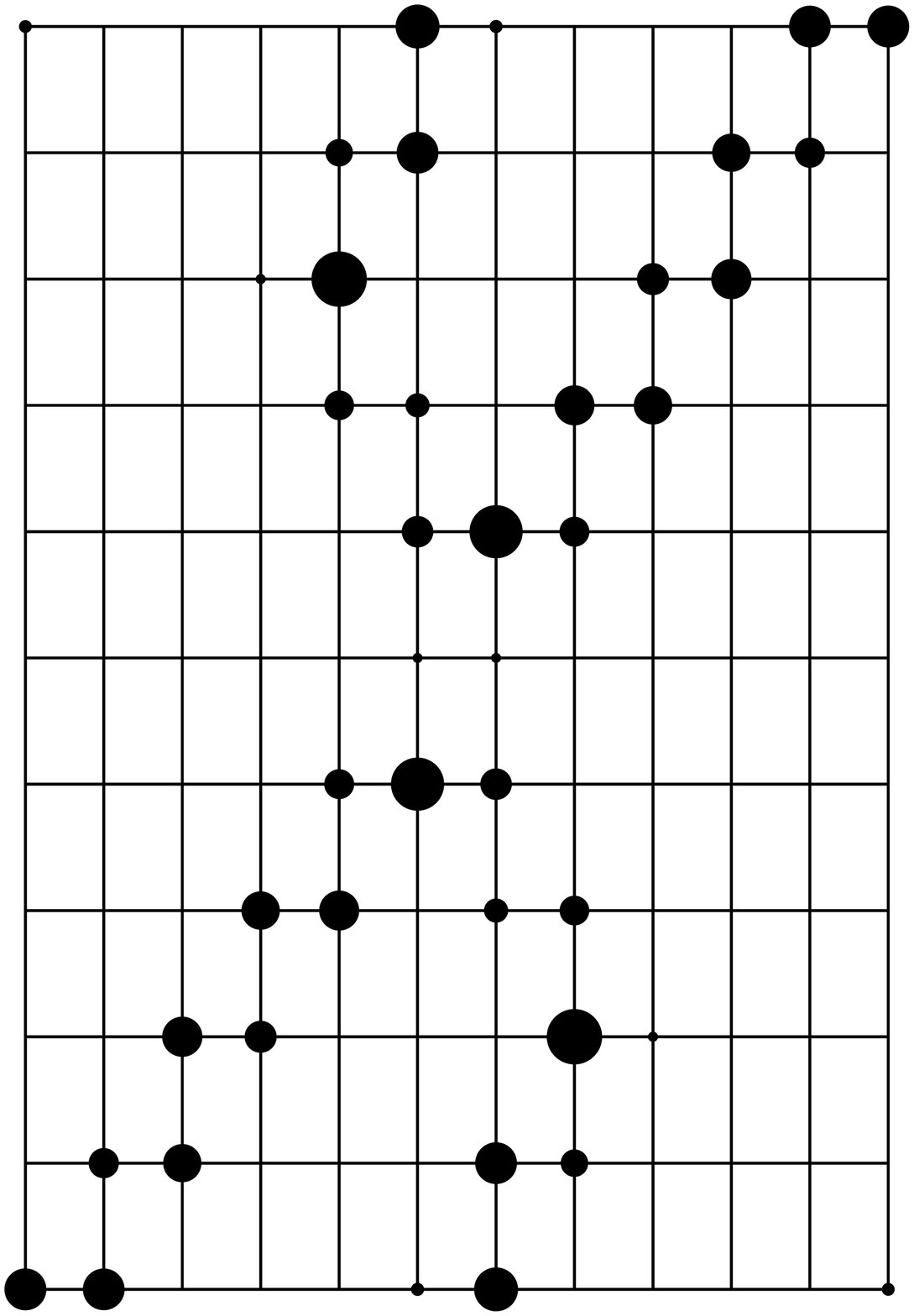,height=5.25cm,angle=0}}
\medskip
{\centerline{(a) \qquad\qquad\qquad\qquad\qquad\qquad (b)}}
\medskip
\caption{Ground-state charge distribution for $t$-$U$-$J$ model on a 
$12 \times 11$ cluster with periodic BCs, for 12 holes ($x = 1/11$), 
$U = 16$, $t_x = t_y = - 1$, and $J_x = J_y = 0.5$. Initial configuration 
with upper left and lower right spins (a) parallel and (b) opposite. 
Charge scale as in Fig.~3. Spin distributions (not shown) are AF with 
$\pi$ phase shifts across all domain walls. }
\end{figure}

Variations of the $t$- and $J$-anisotropies on the two states in Fig.~15 
have little visible effect on the charge configurations even at the 22\% 
level, although in Fig.~15(b) there is noticeable charge redistribution 
to a more ``vertical'' local stripe orientation. Thus we revert 
to analyzing the energies of the ground states, which are presented 
in Tables VI and VIII in the same format as in Tables IV and V. It is 
immediately clear that kinetic and magnetic energies are of the same order 
again in the regime around $U = 16$, and it becomes meaningful to consider 
their combination $E_{\rm tot}$. The next striking feature of the data in 
Table VI is the remarkably small change in the magnetic energy $E_m$ as a 
function of $J_x / J_y$: despite the fact that the system appears to contain 
a stripe, and has no frustrated bonds at the boundaries, variation of $E_m$ 
with $J_x / J_y$ is in fact weaker than with $t_x / t_y$. Another interesting 
property is that the kinetic energy decreases for both types of hopping 
anisotropy, a result which would not be expected for a vertical stripe. 

\medskip
\begin{table}[t!]
\narrowtext
\caption{Kinetic (upper figure) and magnetic (lower) energy components 
per site for $t$-$J$ model on a $12 \times 11$ cluster with 12 holes, 
periodic BCs, $U = 16$, and a starting configuration of parallel spins 
on opposite corners; charge distribution in Fig.~15(a). }
\medskip
\begin{tabular}[hp]{c|c|c|c}
& $t_x = - 0.9$ & $t_x = - 1 $ & $t_x = - 1.1$ \\
& $t_y = - 1.1$ & $t_y = - 1 $ & $t_y = - 0.9$ \\ \hline
$J_x = 0.45$ $\,$ & $\,$ -0.56846 $\,$ & $\,$ -0.55757 $\,$ & $\,$ 
-0.56347 \\
$J_y = 0.55$ $\,$ & $\,$ -0.40620 $\,$ & $\,$ -0.40098 $\,$ & $\,$ 
-0.39812 \\ \hline
$J_x = 0.5$ $\,$ & $\,$ -0.56789 $\,$ & $\,$ -0.55738 $\,$ & $\,$ 
-0.56349 \\
$J_y = 0.5$ $\,$ & $\,$ -0.40712 $\,$ & $\,$ -0.40160 $\,$ & $\,$ 
-0.39848 \\ \hline
$J_x = 0.55$ $\,$ & $\,$ -0.56733 $\,$ & $\,$ -0.55718 $\,$ & $\,$ 
-0.56350 \\
$J_y = 0.45$ $\,$ & $\,$ -0.40809 $\,$ & $\,$ -0.40228 $\,$ & $\,$ 
-0.39888  
\end{tabular}  
\end{table}

The explanation for both of these features is contained in Table VII, 
which shows the components of $E_K$ and $E_m$ for the ${\hat x}$- and 
${\hat y}$-directions separately. The magnetic energies (lower figures) 
vary linearly with $J_x$ and $J_y$, with their sum (Table VI) almost a 
constant quantity. The kinetic energies change very strongly with $t_x$ 
and $t_y$, as already observed in Table I, but again their sum changes 
only weakly. It is clear from the figures that the stripe in 
Fig.~15(a) is almost perfectly diagonal in character. The bond-counting 
argument for effects of $J$ on stripe orientation give no preference in the 
case of a diagonal stripe, and show how well this picture works even for a 
system with a broad charge distribution far from the idealized stripe. 
Changes in magnetic energy due to contributions from the hopping term 
are on the 1\% level. The energy gain for both anisotropy directions also 
requires a predominantly diagonal nature, although the minor preference 
for strong $t_y$ indicates that the stripe does preserve a small amount 
of (1,0) character (less than 5\% compared to the changes in $x$- and 
$y$-components). DMRG studies of the $t$-$J$ model also show a significant 
presence of diagonal stripes, and their near-degeneracy in energy with 
vertically oriented domain walls.\cite{rwsl}

\medskip
\begin{table}[t!]
\narrowtext
\caption{Energy components $\langle E_K^x \rangle$, $\langle E_K^y 
\rangle$, $\langle E_m^x \rangle$, and $\langle E_m^y \rangle$
per site for $t$-$J$ model on a $12 \times 11$ cluster with 12 holes, 
periodic BCs, and $U = 16$. }
\medskip
\begin{tabular}[hp]{c|c|c|c}
& $t_x = - 0.9$ & $t_x = - 1 $ & $t_x = - 1.1$ \\
& $t_y = - 1.1$ & $t_y = - 1 $ & $t_y = - 0.9$ \\ \hline
& $\,$ -0.20022 $\,$ & $\,$ -0.26859 $\,$ & $\,$ -0.33180 \\ 
$J_x = 0.45$ $\,$ & $\,$ -0.36824 $\,$ & $\,$ -0.28898 $\,$ & $\,$ 
-0.23167 \\
$J_y = 0.55$ $\,$ & $\,$ -0.18584 $\,$ & $\,$ -0.18201 $\,$ & $\,$ 
-0.17987 \\ 
& $\,$ -0.22036 $\,$ & $\,$ -0.21896 $\,$ & $\,$ -0.21825 \\ \hline
& $\,$ -0.20045 $\,$ & $\,$ -0.26917 $\,$ & $\,$ -0.33278 \\ 
$J_x = 0.5$ $\,$ & $\,$ -0.36745 $\,$ & $\,$ -0.28820 $\,$ & $\,$ 
-0.23070 \\
$J_y = 0.5$ $\,$ & $\,$ -0.20678 $\,$ & $\,$ -0.20255 $\,$ & $\,$ 
-0.20010 \\ 
& $\,$ -0.20034 $\,$ & $\,$ -0.19905 $\,$ & $\,$ -0.19838 \\ \hline
& $\,$ -0.20066 $\,$ & $\,$ -0.26975 $\,$ & $\,$ -0.33376 \\ 
$J_x = 0.55$ $\,$ & $\,$ -0.36667 $\,$ & $\,$ -0.28743 $\,$ & $\,$ 
-0.22973 \\
$J_y = 0.45$ $\,$ & $\,$ -0.22778 $\,$ & $\,$ -0.22313 $\,$ & $\,$ 
-0.22037 \\ 
& $\,$ -0.18031 $\,$ & $\,$ -0.17914 $\,$ & $\,$ -0.17852 
\end{tabular}  
\end{table}

This result is rather surprising in view of the properties observed earlier 
in the Hubbard model, and invite speculation on the relationship of $t$-$U$ 
and $t$-$U$-$J$ models. A discussion of the energetics is incomplete without 
noting that on the same scale as in Tables VI-VIII the double-occupancy 
energy for $U = 16$ is approximately 0.2 in every case, so is a quantity 
which may not be neglected. By contrast, in the previous subsection the 
double-occupancy energy for $U = 1000$ was on the order of 0.0036 in each 
case. 

\medskip
\begin{table}[t!]
\narrowtext
\caption{Kinetic (upper figure) and magnetic (lower) energy components 
per site for $t$-$J$ model on a $12 \times 11$ cluster with 12 holes, 
periodic BCs, $U = 16$, and a starting configuration of antiparallel 
spins on opposite corners; charge distribution in Fig.~15(b). }
\medskip
\begin{tabular}[hp]{c|c|c|c}
& $t_x = - 0.9$ & $t_x = - 1 $ & $t_x = - 1.1$ \\
& $t_y = - 1.1$ & $t_y = - 1 $ & $t_y = - 0.9$ \\ \hline
$J_x = 0.45$ $\,$ & $\,$ -0.55432 $\,$ & $\,$ -0.56027 $\,$ & $\,$ 
-0.56816 \\
$J_y = 0.55$ $\,$ & $\,$ -0.35716 $\,$ & $\,$ -0.38007 $\,$ & $\,$ 
-0.38864 \\ \hline
$J_x = 0.5$ $\,$ & $\,$ -0.55126 $\,$ & $\,$ -0.55928 $\,$ & $\,$ 
-0.56619 \\
$J_y = 0.5$ $\,$ & $\,$ -0.35322 $\,$ & $\,$ -0.37394 $\,$ & $\,$ 
-0.38890 \\ \hline
$J_x = 0.55$ $\,$ & $\,$ -0.55414 $\,$ & $\,$ -0.55929 $\,$ & $\,$ 
-0.56597 \\
$J_y = 0.45$ $\,$ & $\,$ -0.35327 $\,$ & $\,$ -0.37393 $\,$ & $\,$ 
-0.38927  
\end{tabular}  
\end{table}

Table VII shows the kinetic and magnetic energies under anisotropy changes 
of 22\% in $t$ and $J$ for the charge configuration of Fig.~15(b). Once 
again, the values are comparable, and they are also very close indeed to 
the figures in Table VI for a very different charge configuration; in fact 
it is the magnetic energy which would drive a preference for the broad, 
meandering stripe of Fig.~15(a). In Fig.~15(b) there appears to be a 
higher net alignment of the domain walls along the ${\hat y}$-direction, 
and this may be taken to explain the preference shown by the kinetic energy 
for stronger $t_x$. Here we notice a rather strong effect of the hopping 
on the magnetic energy, for which the bond-counting argument also provides 
an explanation. In narrow stripes, a diagonal orientation breaks four bonds 
per site while a vertical one breaks three, and thus a hopping anisotropy 
(of either direction) which pushes a diagonal stripe towards a vertical one 
to gain transverse hopping energy will also lower the magnetic energy. 
That the net magnetic energy is lower for stronger $J_y$ is also consistent 
with the inferred (0,1) character. The fact that the numbers in the lower 
two rows match to very high accuracy is not an error, as shown by the fact 
that the separate $x$- and $y$-components of the relevant energies (not 
presented here) have the same 20\% and 40\% differences as in Table VII, and 
rather reflects the delicate nature of the energy balance for the weakly
oriented domain-wall structure. 

We have also considered the same calculations with different values of 
$U (= 50)$ and of $J (= 0.25)$, and find qualitatively similar results. 
Changing the filling $x$ also leaves the general conformation unaltered, 
although the domain wall loops in Fig.~15(b) become shorter. The 
analysis for an $11 \times 11$ system returns the same spin configurations 
as in Ref.~\onlinecite{rvr}, although we find the 12-hole domain line to 
be rather broad. For $12 \times 12$ arrays the domain lines tend to close,
to avoid disturbing BCs which are already satisfied. 

We close this section with a summary of the robust conclusions which may 
be drawn from our studies of the $t$-$J$ model. Without considering 
anisotropy, the superexchange term favors polarons over corrals. The 
presence of $J$ also leads to a form of phase separation, namely the 
formation of charge clumps near the edges of an open system. As a result, 
in the HF approximation, only systems with rather small values of $J$ 
may show formation of domain walls, whose alignment tends to be 
diagonal. This tendency is particularly strong in the $t$-$U$-$J$ model 
with intermediate values of $U$. Hopping anisotropy $t_x \ne t_y$ promotes 
uniform stripes normal to the strong hopping direction, as in the Hubbard 
model. There is a parameter regime of intermediate $J$ which favors 
small polarons, but no clear sign of the non-uniform, in-phase stripes 
observed for the Hubbard model. Superexchange anisotropy $J_x \ne 
J_y$ favors alignment of uniform stripes in the direction of stronger 
exchange, and thus competes with a hopping anisotropy of the same 
orientation. While quantitative results are hard to obtain, energetic 
considerations suggest that the relative effects of $t$ anisotropy are 
stronger. For the boundary region of $(t,J,x)$ parameter space between 
polaronic states and phase separation, we find that anisotropies in $t$ 
and $J$ act to stabilize coherent structures.

\section{Discussion and Conclusion}

In the previous sections we have detailed at length the many degrees of 
freedom allowed by the real-space HF approach, and the ways in which we 
use and interpret these to draw meaningful conclusions within this 
framework. We have also reviewed the experimental situation, focusing 
on those parts we are able to address, and have compared our results with 
existing theoretical literature. In this section we give a brief discussion 
of the more general context of our analysis. 

In Sec. II we noted that the LTT distortion of the LSCO system provides 
anisotropies of 1-3\% in the hopping and superexchange parameters.\cite{rksw} 
Our investigations in Secs. IV and V required anisotropies on the order of 
10-20\%, and these were not always sufficient to create 
a 1d state. We appeal here to the quantitative insufficiencies of the HF 
technique to argue that our results indeed suggest an important role for 
anisotropies, as suggested by the comparison of the hopping energy difference 
with other key energy scales. The superconducting $T_c$, and the ordering 
temperature $T_{\rm ch}$ of stripe formation, are both around 40K in LSCO. 
One of the very few indisputable results from the extensive studies of 
cuprate models is that there exist many competing candidate ground states 
whose energies are remarkably similar, and it seems not unreasonable that 
small anisotropies may play a significant role in selecting among these. 
In addition to the more detailed energetic considerations and analytical 
or numerical approaches which we mention below, a more accurate assessment 
of anisotropy effects would also require moving from a one-band model to 
the three-band case.\cite{rzg} This step is also a prerequisite for 
discussing at the microscopic level the issue of bond- or site-centered 
stripes,\cite{rwsl,rzeah} which is one we have not mentioned hitherto. All 
of the charge structures we find by unrestricted HF for the one-band model, 
with and without anisotropy, are site-centered, and we do not regard the 
model as appropriate for addressing this question. 

The connection between the lattice structure and anisotropies in the 
electronic system suggests the importance of the phonon degrees of 
freedom. While dynamical phonons may play a significant role around 
the structural phase transition, the static distortion requires a 
slow phonon mode which becomes soft. Although static, averaged 
phonon variables have been included in some HF analyses,\cite{rgro} 
we remain unaware of any studies where the coupled phonon modes which 
are introduced have allowed the possibility of a spontaneous anisotropy.
As noted in Sec. II, we have studied only a model with fixed anisotropy 
set by the lattice, and considered the energetics of the electronic 
system. When the structural transition is in close proximity to the 
electronic one, the need for a coupled model becomes apparent. The 
relevance of lattice energies for the electronic system may also be 
clarified by detailed experimental analysis of the energetics of the 
structural transition. 

Other transition-metal oxide systems offer a clear example of the 
coupling of lattice and electronic degrees of freedom. The first 
observations of charge-ordering in stripe-like structures were made 
on nickelates, where the stripes are found to be diagonal (aligned 
along (1,1)). Detailed characterization of the ordering transition 
suggests once again a connection to the orthorhombic lattice structure 
arising from octahedral tilting at dopings $x < 0.2$.\cite{rsbtls} 
Above this doping, charge ordering is a coupled structural and 
electronic event, which occurs at a single temperature, $T_d$, and 
this behavior persists to very high dopings (La$_{2-x}$Sr$_x$NiO$_4$ 
with $x \le 0.5$).\cite{rykktkt} In this case the energy change 
of the electronic system on adopting an anisotropic structure must 
be balanced directly against the distortion energy of the lattice, and 
theoretical estimates of the competing contributions become more involved. 
In manganite systems, the most common structural deformation is a 
Jahn-Teller distortion, which directly affects the electronic system by 
lifting the $e_g$ orbital degeneracy.\cite{rcfg} Such distortions usually 
alternate between sites (``AF''), and experimentally they appear to 
influence the nature of the ground state primarily at low hole doping. 
However, observation of stripe-like structures in manganite systems\cite{rmcc} 
has not yet been related unambiguously to lattice distortions, and may 
lie in the competition of next-neighbor repulsion $V$ with anisotropic 
conduction-electron hopping $t$.

With regard to methodology, the Hartree-Fock approach is generally 
understood to be ``crude but effective''. It is capable of finding the 
relevant ground states of the system, and allows one to draw many 
qualitative conclusions about their possible nature and order. However, 
the approximations involved in the breaking of symmetries and neglect 
of quantum fluctuations are sufficiently strong that the parameter ranges 
where different features appear, {\it i.e.} the quantitative aspect, may 
not be trusted. In addition, with periodic BCs, the self-consistency 
procedure has difficulties converging adequately due to the large choice 
of (translationally) degenerate ground states. A possible solution to 
both problems is offered by the Configuration Interaction (CI) 
method.\cite{rf,rlglv2,rbj} In this approach, the ground state is 
constructed as a (biased) linear combination of HF solutions. This 
technique, which has already been applied to certain aspects of the 
isotropic Hubbard model,\cite{rlglv2,rlglv1} can be shown to give a 
systematic description of quantum fluctuation and tunneling corrections 
to the HF solution, and of the quantum dynamics of the excitations.

While CI represents a significant improvement to HF, there remain many other 
techniques which could be applied to address the question of anisotropy 
effects at a more quantitative level. As listed in Sec. I, these include 
effective models of domain walls, the slave-boson representation, DMFT 
and DMRG, and we await with interest the results of subsequent analyses. 
We note finally that similar or adapted HF techniques may also be expanded 
to explore situations beyond the Hubbard and $t$-$J$ models. In the 
framework of more general $t$-$J$-$U$-$V$ models one may seek finite, 
nearest-neighbor pairing\cite{rmobb} and spin-flip\cite{rbj} order 
parameters, with a view to discussing superconductivity and flux phases 
in charge and spin. The latter possibilities have been of considerable 
recent interest in both single- and multi-band models. 

\medskip
\begin{figure}[t!]
\centerline{\psfig{figure=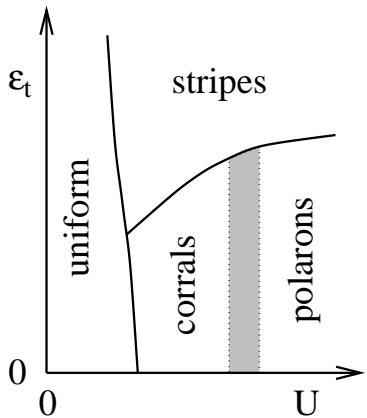,height=5.5cm,angle=0}}
\medskip
\caption{Schematic phase diagram of HF solutions for the Hubbard model as a 
function of $U$ and hopping anisotropy $\epsilon_t$, for dopings $0.06 < x 
< 0.2$. The ``uniform'' region is a paramagnetic metal. The nature of the 
stripe phases depends on $x$, as discussed in Sec. IV. Solid lines are 
true phase transitions, while the character of charge-inhomogeneous 
solutions at large $U$ and small $\epsilon_t$ changes continuously. }
\end{figure}

In conclusion, certain cuprate systems show a breaking of 4-fold $(x,y)$ 
symmetry in the CuO$_2$ plane which is closely connected to the appearance 
of static stripes. Proceeding from this experimental motivation we have 
considered Hubbard and $t$-$J$ models with hopping and superexchange 
anisotropy. The real-space Hartree-Fock approach offers many possibilities 
by which to gain insight into the nature of the solutions. In the isotropic 
Hubbard model we find predominantly closed, diagonal domain-wall loops 
(corrals), and polarons at larger $U$. Anisotropy quite generally brings 
a gain in electronic energy, and introduces vertical stripes, which are 
stabilized by hopping between different charge densities. The qualitative 
situation is summarized in Fig.~16, which without distinguishing between 
types of stripe is applicable for the doping range of 
physical interest. The dominant, uniformly ``filled'' stripes are oriented 
perpendicular to the direction of strong hopping, have antiphase nature, 
and show in their ground-state energy the very strong influence of 
transverse hopping. For certain hole densities a non-uniform stripe 
appears, which may be the HF analogue of 
the ``half-filled'' stripes seen in experiment, and has longitudinal 
orientation and in-phase character. The isotropic $t$-$J$ model in the 
HF approximation favors polaron formation and shows a tendency to phase 
separation with increasing $J$. Anisotropies in $t$ favor transverse 
stripe formation, while anisotropies in $J$ reinforce longitudinal 
orientation of stable stripes, but also act to promote phase separation. 
We conclude that stripes cannot be considered in isolation from the 
question of lattice-induced anisotropies.

\section*{Acknowledgements}

We are grateful to C. Morais Smith, A. M. Ole\'s, T. M. Rice, D. J. 
Scalapino, M. Vojta, R. Werner, and S. R. White for invaluable discussions. 
This work was supported by the Deutsche Forschungsgemeinschaft through SFB 
484.

\end{document}